\documentclass[lettersize,journal]{IEEEtran}
\usepackage{amsmath,amsfonts}
\usepackage{algorithm, algpseudocode}
\usepackage{array}
\usepackage[caption=false,font=normalsize,labelfont=sf,textfont=sf]{subfig}
\usepackage{textcomp}
\usepackage{stfloats}
\usepackage{url}
\usepackage{verbatim}
\usepackage{graphicx}
\usepackage{cite}
\usepackage{amsmath,amssymb,amsfonts,bm}
\usepackage{xcolor}
\usepackage{dsfont}
\usepackage{multirow}
\usepackage{booktabs}
\usepackage{makecell}
 \usepackage{tikz}
 \usepackage{subcaption} 
 \usepackage[hidelinks]{hyperref}
 \usepackage{tabularx}
 \usetikzlibrary{arrows.meta,positioning,calc,shapes.geometric,shapes.symbols,matrix}
 \tikzset{
 	updstar/.style={
 		star, star points=5, star point ratio=2.0,
 		draw, line width=0.35pt,
 		inner sep=0pt, outer sep=0pt, 
 		transform shape           
 	}
 }
\newtheorem{assumption}{Assumption}
\newtheorem{theorem}{Theorem}

\def\changeBibColor#1{%
	\in@{#1}{}
	\ifin@\color{blue}\else\normalcolor\fi
}

\def\black {\color{black}}
\definecolor{myred}{HTML}{C00000}

\begin{document}

\title{RadioTrace: Transmitter-Aware Diffusion for Radio Map Estimation without Deployment-Time Fine-Tuning}

\author{\IEEEauthorblockN{Liu Yang, Qiang Li, Zhuo Cao, Weijie Xiong, Guomin Sun and Jingran Lin}
	\thanks{A preliminary version of this work was presented at 35th IEEE International Workshop on Machine Learning for Signal Processing (MLSP)~\cite{yang2025radiotrace}.
		
		This work was supported by the National Natural Science Foundation of China under Grant 62301115.
		
		L. Yang, Z. Cao and W. Xiong   are with
		School of Information and Communication  Engineering, University of Electronic Science and Technology of China, Chengdu, China, 611731.  
		
		Q. Li, G. Sun and J. Lin are with School of Information and Communication  Engineering, University of Electronic Science and Technology of China, Chengdu, China, 611731, and also with Tianfu Jiangxi Laboratory, Chengdu, China, 641419.

		Q. Li is the corresponding author. E-mail: lq@uestc.edu.cn.}}



\maketitle

\begin{abstract}

Radio map (RM) estimation aims to reconstruct the spatial distribution of wireless signal characteristics, such as received signal strength (RSS), from sparse measurements, a task that is critical for spectrum management, interference mitigation, and localization in modern wireless networks. Traditional approaches, including interpolation and deep learning, either struggle to capture complex propagation effects or require large-scale retraining for each new sampling pattern, which limits their generalization. {\black More recently, prior-based methods have combined pre-trained generative models with measurements to reduce the need for deployment-time model fine-tuning, but they typically treat the prior as a simple regularizer and lack explicit transmitter-aware integration.}
{\black In this paper, we propose RadioTrace, a novel RM estimation framework without deployment-time fine-tuning that tightly integrates sparse RSS measurements with a frozen pre-trained diffusion prior.} RadioTrace incorporates transmitter (Tx) location estimation directly into the denoising loop, iteratively refining Tx coordinates based on reconstruction quality to guide the generative process. To further enhance robustness, we introduce a propagation-guided K-means initialization that mitigates poor local minima in the Tx update and provides a geometry-consistent starting point. {\black Moreover, we provide a stochastic stability analysis for the Tx-coordinate refinement component, showing that the Tx update remains stable under perturbations induced by diffusion sampling and Tx-map relaxation.}
Extensive experiments demonstrate that RadioTrace achieves competitive performance with state-of-the-art learning-based methods under random sampling, and maintains strong reconstruction quality under restricted-area sampling, highlighting its adaptability, robustness, and practical relevance.
\end{abstract}


\section{Introduction}

\begin{figure}[tb]
	
	\centering
	\centerline{\includegraphics[width=9cm]{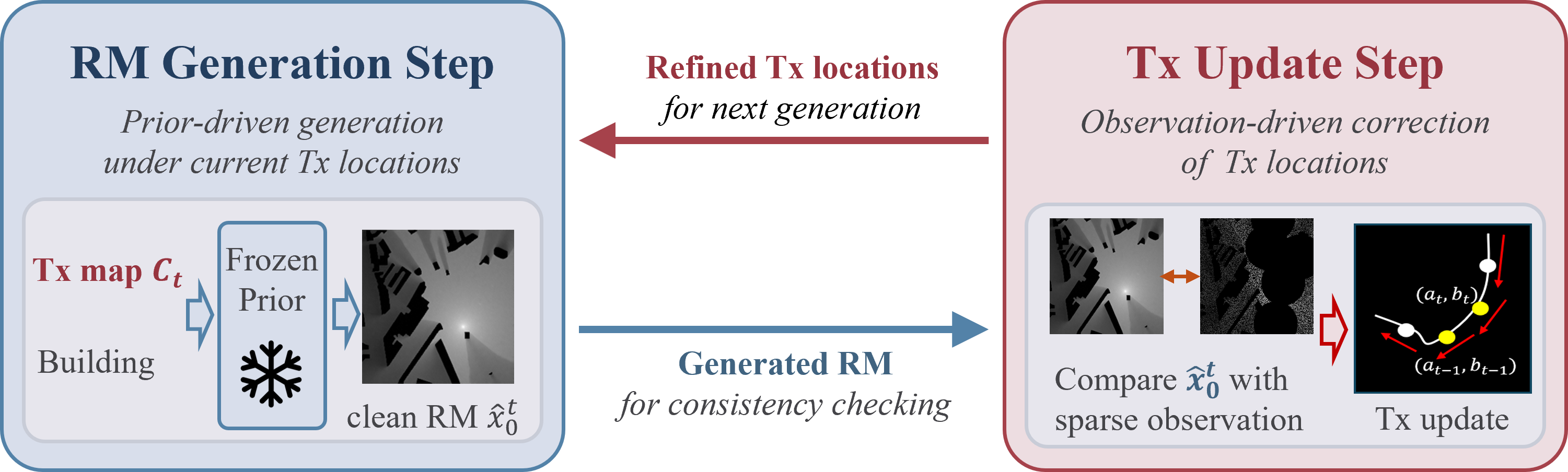}}
	
	\caption{Conceptual illustration of the loop interaction between the RM generation step and the Tx update step in RadioTrace.}
	\label{fig:relation}
\end{figure}

{\black
	Radio maps (RMs) characterize the spatial distribution of wireless signal attributes, such as received signal strength (RSS), signal-to-noise ratio (SNR), and power spectral density (PSD). They provide radio-environment information for spectrum management, interference monitoring, base-station planning, localization, and adaptive network optimization. RMs are also relevant to wireless digital twins and integrated sensing and communication (ISAC), where propagation knowledge supports both communication and environmental awareness~\cite{bi2019eng,chen2024optimal,zeng2024tutorial,fu2025ckmdiff,zhao2025imnet}. This paper focuses on RM reconstruction from limited RSS measurements and investigates how prior propagation knowledge can be incorporated into this task.
	
	Existing RM reconstruction studies can be roughly divided into two lines. The first line is measurement-free RM generation, which synthesizes complete RMs from environmental priors, such as building layouts, material assumptions, and transmitter (Tx) locations. Classical ray-tracing methods provide physically interpretable predictions based on geometric propagation models~\cite{yun2015ray}. Recent learning-based methods, including U-Net-based generators~\cite{Levie2021RadioUNet} and diffusion models such as RadioDiff~\cite{Wang2024RadioDiff}, improve generation efficiency by learning propagation patterns from data. Although these methods do not directly solve sparse-measurement-based RM estimation, they provide useful generative priors for radio propagation modeling.
	
	The second line directly estimates RMs from sparse RSS measurements. Classical interpolation methods, such as Kriging~\cite{Lee2012Voronoi,Romero2022Radio}, exploit spatial correlation among samples but are limited in complex environments with blockage, shadowing, and multipath effects. Supervised deep models, including auto-encoder-based methods~\cite{Teganya2022DeepCA}, generative adversarial networks~\cite{Zhang2023RMEGAN,zhang2024fast}, and diffusion-based estimators~\cite{jia2025rmdm,luo2025denoising}, learn nonlinear mappings from sparse observations to complete RMs. More recently, inverse reconstruction methods have been explored to reduce the need for deployment-time fine-tuning. For example, LaPnP combines measurement consistency with latent-domain plug-and-play denoising~\cite{Xu2025Radio}, while RadioDiff-Inverse formulates sparse-measurement RM reconstruction as a Bayesian inverse problem with a diffusion prior~\cite{wang2025radiodiff}.

Despite these advances, existing methods still face several limitations in practical sparse-measurement RM estimation. 
First, measurement-free generative models can provide strong propagation priors, but they often rely on source-related conditions, especially accurate Tx locations. 
In civilian spectrum monitoring with covert unauthorized Tx or in military scenarios involving hostile emitters, such information may be unavailable or deliberately concealed. 
Second, supervised sparse-measurement estimators typically require large-scale paired training data and may generalize poorly when the deployment sampling pattern differs from the training distribution. 
In practice, access restrictions, sensing budgets, and task-specific coverage requirements often lead to nonuniform or restricted-area sampling, under which additional fine-tuning may be required.
Third, although prior-based inverse reconstruction methods couple denoising or generative priors with sparse measurements, their measurement consistency is mainly imposed at the RM level. For instance, LaPnP refines the reconstructed RM estimate within a PnP reconstruction loop, whereas RadioDiff-Inverse corrects posterior RM samples during reverse diffusion. These methods use sparse measurements to improve the agreement between the reconstructed RM and the observed RSS values in sampled cells. However, the underlying physical factors that govern radio-map formation, such as the propagation geometry, are usually not explicitly modeled or updated during the reconstruction process. 
This motivates a key question: how can sparse measurements be used not only to enforce RM-level consistency, but also to infer and embed the relevant physical information into the generative RM reconstruction process?

To fill in this gap, we explicitly incorporate Tx locations into RM estimation, and propose RadioTrace, a Tx-aware diffusion framework for RM estimation without deployment-time fine-tuning. RadioTrace uses a frozen pre-trained diffusion model as the RM generation prior, while inferring the unknown Tx locations from sparse RSS measurements during reverse diffusion. Specifically, the current Tx estimates condition RM generation, and the discrepancy between the generated RM and sparse observations is back-propagated to refine these estimates. The refined Tx locations are then fed back into subsequent denoising steps, forming a closed-loop diffusion--geometry inference process, as illustrated in Fig.~\ref{fig:relation}.

\begin{table}[t] \black
	\centering
	\caption{
		Positioning of RadioTrace relative to prior-based inverse RM reconstruction methods.
	}
	\label{tab:method_positioning}
	\footnotesize
	\renewcommand{\arraystretch}{1.2}
	\begin{tabularx}{\linewidth}{@{}p{0.23\linewidth}X X@{}}
		\toprule
		\textbf{Method}
		& \textbf{Modeling level}
		& \textbf{Updated variable} \\
		\midrule
		\textbf{LaPnP~\cite{Xu2025Radio}}
		& RM level
		& Reconstructed RM estimate \\
		\midrule
		\textbf{RadioDiff-Inverse~\cite{wang2025radiodiff}}
		& RM level
		& Posterior RM samples \\
		\midrule
		\textbf{RadioTrace}
		& Tx-condition level
		& Tx locations + RM \\
		\bottomrule
	\end{tabularx}
\end{table}

Table~\ref{tab:method_positioning} summarizes the modeling-level
distinction between RadioTrace and representative prior-based inverse
RM estimation methods. Existing methods mainly operate at the RM
level by updating reconstructed RM estimates or posterior RM samples,
whereas RadioTrace refines the Tx locations that condition subsequent
RM generation.

	The main contributions of this paper are summarized as follows: 
	\begin{itemize} 
		\item From a modeling perspective, we develop a physics-driven formulation that jointly models the RM and Tx levels, treating unknown Tx locations as physical latent variables and casting RM estimation as a joint RM–Tx recovery problem.
		\item From an algorithmic perspective, we develop RadioTrace, an alternating inference framework that embeds Tx estimation into reverse denoising, enabling deep integration between sparse measurements and the diffusion prior.
		\item From a theoretical perspective, we establish the stochastic stability of Tx-coordinate refinement under reverse-diffusion-induced time variation, showing that it asymptotically enters a stationary regime under some conditions.
\end{itemize}}

\section{Problem Formulation}
Consider a rectangular region of size $w \times h$ containing $R$ Tx. These Tx simultaneously emit signals that propagate through the environment, resulting in a spatially varying RSS distribution over the area. Let ${\bm X} \in \mathbb{R}^{w \times h}$ denote the complete RM over the area of interest, where each element ${\bm X}(m, n)$ represents the received signal strength at location $(m, n)$ (in dBm). However, this complete map is not directly accessible in practice. Instead, we can only obtain partial information by deploying a limited number of sensors at specific locations to perform signal measurements. 
Each sensor captures the RSS over a small spatial grid cell. This sparse sampling process can be described by a binary mask $\mathcal{M} \in \{0,1\}^{w \times h}$, where $\mathcal{M}(m, n) = 1$ indicates that the grid point $(m, n)$ has been measured, and 0 otherwise. The observed measurements ${\bm O} \in \mathbb{R}^{w \times h}$ are therefore obtained as:
\begin{equation}
	{\bm O} = \mathcal{M} \odot {\bm X},
\end{equation}
where $\odot$ represents the element-wise product. The goal of RM estimation is to recover the complete  radio maps  ${\bm X}$ from sparse RSS measurements $\bm O$, the sampling mask $\cal M$ and available environmental information.

\begin{figure*}[htb]
	
	\centering
	\centerline{\includegraphics[width=18cm]{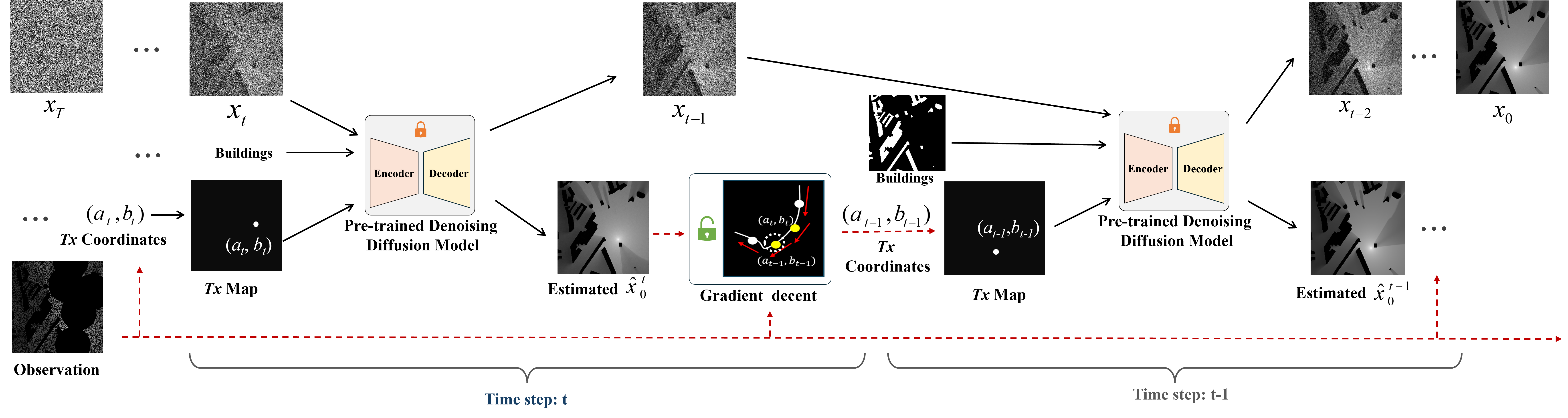}}
	
	\caption{Overview of the proposed RadioTrace (\textbf{black arrow}: RM generation step, \textbf{\color{myred}red arrow}: Tx update step).}
	\label{fig:overview}
\end{figure*}

\section{RadioTrace}
In this section, we present the proposed RadioTrace framework. Section~\ref{subsec:overview} outlines the main idea, while Sections~\ref{subsec:rmgen}–\ref{sec:pgkmeans} describe the detailed implementation.


\subsection{Overview}
\label{subsec:overview}

{\black 
	
	As illustrated in Fig.~\ref{fig:relation} and detailed in Fig.~\ref{fig:overview}, RadioTrace embeds the RM-generation and Tx-coordinate-refinement loop into a reverse diffusion process.} Starting from a Gaussian initialization ${\bm x}_T\!\sim\!\mathcal{N}(\mathbf{0},{\bm I})$, the method proceeds through $T$ denoising steps. At each reverse step $t$, RadioTrace alternates between two steps:
	
	\begin{itemize}
			\item \textbf{RM Generation Step}: Given the current state ${\bm x}_t$, the iteration index $t$, and a binary Tx map ${\bm C}_t$, the fixed generator produces a clean estimate $\hat{\bm x}_0^t$ and the denoised RM $\bm x_{t-1}$ for the next step $t-1$.
			
			\item \textbf{Tx Update Step}: The Tx locations are updated via gradient descent to minimize the reconstruction error between $\hat{\bm x}_0^t$ and the observation $\bm O$.

	\end{itemize}
	By repeating the two steps throughout the reverse diffusion process, RadioTrace progressively recovers both the RM and the Tx locations. In the following subsections, we will explain these two steps in detail.

\subsection{RM Generation Step}
\label{subsec:rmgen}

At reverse step $t$, the RM generation step takes the noisy state ${\bm x}_t\in\mathbb{R}^{w\times h}$, the building map ${\bm B}$ and the \emph{binary} Tx map ${\bm C}_t\in\{0,1\}^{w\times h}$ (where ${\bm C}_t(m,n)=1$ if and only if a Tx is present at $(m,n)$), and invokes a measurement-free RM generator (diffusion prior) $f_\theta$ whose parameters are trained offline and remain fixed throughout RadioTrace. The generator inputs $({\bm x}_t, t, {\bm C}_t)$, produces a clean estimate $\hat{\bm x}_0^t$, and parameterizes the conditional distribution of ${\bm x}_{t-1}$, from which we sample to continue the reverse trajectory. In this step, ${\bm C}_t$ is treated as a given conditioning signal and its representation and subsequent refinement are deferred to Section~\ref{subsec:txupdate}. 

\begin{figure}[tbp]
	\centering
	\includegraphics[width=8.5cm]{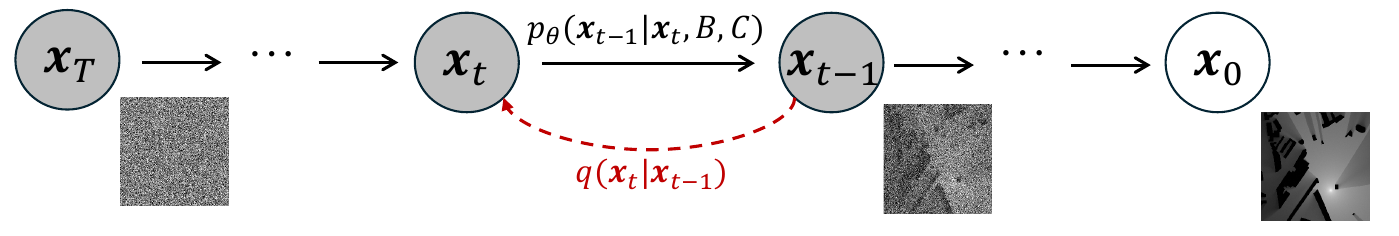}
	\caption{Measurement-free RM generation with diffusion models.}
	\label{fig:ddpm}
\end{figure}

As illustrated in Fig.~\ref{fig:ddpm}, the black arrows depict the reverse sampling path used at inference.
Operationally, this step performs a single reverse-diffusion update conditioned on $({\bm x}_t,t,{\bm C}_t)$. The generator predicts the injected noise $\epsilon_\theta$ and yields
\begin{equation}\label{eq:x0t}
	\hat{\bm x}_0^t
	=\frac{1}{\sqrt{\bar\alpha_t}}\Big({\bm x}_t-\sqrt{1-\bar\alpha_t}\,\epsilon_\theta({\bm x}_t,t,{\bm B},{\bm C}_t)\Big),
\end{equation}
where $\bar\alpha_t=\prod_{i=1}^t\alpha_i$ follows a fixed noise schedule. The mean of the next noisy state is
\begin{equation}
	\mu_\theta({\bm x}_{t-1})
	=\frac{\sqrt{\alpha_t}(1-\bar\alpha_{t-1})\,{\bm x}_t+(1-\alpha_t)\sqrt{\bar\alpha_{t-1}}\,\hat{\bm x}_0^t}{1-\bar\alpha_t},
\end{equation}
and we draw
\begin{equation}\label{eq:sample}
	{\bm x}_{t-1}\sim p_\theta(\cdot\mid {\bm x}_t,{\bm B},{\bm C}_t)
	=\mathcal N\!\big({\bm x}_{t-1};\,\mu_\theta({\bm x}_{t-1}),\,\gamma_t{\bm I}\big),
\end{equation}
with variance $\gamma_t=\tfrac{(1-\bar\alpha_{t-1})({1-\alpha_t})}{1-\bar\alpha_t}$. And the next step noisy RM ${\bm x}_{t-1}$ can be sampled from $p_\theta({\bm x}_{t-1} | {\bm x}_t, {\bm C}_t)$.

\medskip
\noindent\textit{Remark.}
{\black The RM generation step relies on a pre-trained diffusion model $f_\theta$. There are various ways to obtain such a model. For instance, one may pre-train $f_\theta$ offline through a measurement-free RM generation task, as in RadioDiff~\cite{Wang2024RadioDiff}. Our framework primarily focuses on how to leverage this pre-trained model.}

%
%

\subsection{Tx Update Step} \label{subsec:txupdate}
The measurement-free diffusion prior employed in the RM generation stage can produce accurate radio maps when provided with a precise binary Tx map \( {\bm C}_t \). However, in RM estimation, \( {\bm C}_t \) is unknown. A straightforward approach is therefore to recover \( {\bm C}_t \) from the sampled RSS by minimizing the masked reconstruction error:
\begin{equation}
	\min_{{\bm C}_t\in\{0,1\}^{w\times h}}
	\ \big\|\,\hat{\bm x}_0^t({\bm C}_t)\odot\mathcal{M}-\bm O\,\big\|_F^2,
	\label{eq:directC}
\end{equation}
where $\|\cdot\|_F$ is the Frobenius norm. However, since ${\bm C}_t$ is high-dimensional and discrete, direct optimization is challenging. To reduce the degrees of freedom while retaining physical interpretability, we represent the Tx by their continuous coordinates \(\Omega_t=\{(a_t^i,b_t^i)\}_{i=1}^R\in\mathbb{R}^{2R}\), and parameterize the binary Tx map $C_t$ through the following relation:
\begin{equation}
	{\bm C}_t(p,q)\triangleq
	\begin{cases}
		1, & \exists\, i\in\{1,\dots,R\}\ \text{s.t. } (p,q)=(u_t^i, v_t^i),\\
		0, & \text{otherwise},
	\end{cases}
	\label{eq:hardC}
\end{equation}
where $(u_t^i, v_t^i)=\mathrm{round}\big(a_t^i,b_t^i\big)$. Thus $\Omega_t$ becomes the actual optimization variables that influence the RM generation through ${\bm C}_t$, which serves as the input to the reverse diffusion process at step $t$. This parameterization
dramatically reduces the variable dimension from $w\times h$ binary  elements to $2R$ continuous ones, significantly improving computational efficiency. 

To  further stabilize the reverse-time refinement, we introduce a best-anchor mechanism and optimize an augmented objective with a proximal regularization term:
\begin{equation}\label{eq:one_loss}
	\mathcal{L}_t(\Omega_t)
	\;=\;
	\big\|\hat{\bm x}_0^t(\Omega_t)\odot\mathcal{M}-\bm O\big\|_F^2
	\;+\;
	\frac{\kappa_t}{2}\,\big\|\Omega_t-\Omega_t^\star\big\|_F^2,
\end{equation}
where  $\Omega_t^\star$ denotes  the best Tx coordinates found up to time   $t$, 
\[
\Omega_t^\star \;\triangleq\; \mathop{\arg\min}_{\Omega_s,s\ge t}\ \big\|\hat{\bm x}_0^s(\Omega_s)\odot\mathcal{M}-\bm O\big\|_F^2,
\]
and $\kappa_t$ is a linearly scheduler defined as
\[
\kappa_t \;=\; \kappa\,\frac{T-t}{T},\ \ \kappa>0.
\]
The quadratic term acts as a proximal pull, shrinking the search region toward the anchor and suppressing oscillations when the loss landscape becomes sharp at low noise levels. The linear schedule ensures that this pull is negligible during early exploratory stages and gradually strengthens as the noise level decreases.
As shown later in Section~\ref{sec:Exp}, the proposed best-anchor mechanism substantially enhances the recovery accuracy of RM estimation.

To minimize $\mathcal{L}_t(\Omega_t)$, we apply momentum-based gradient descent (GD) method to update the coordinates. For each horizontal coordinate $a_t^i$, the update is given by:
\begin{equation}
	v_{t-1} = \beta v_{t} + (1 - \beta) \frac{\partial \mathcal{L}_t}{\partial a_t^i}, \quad a_{t-1}^i = a_t^i - \eta v_{t-1},
\end{equation}
where $\eta > 0$ is the learning rate and $\beta > 0$ is the momentum coefficient. The same update rule is applied to the vertical coordinates $b_t^i$.


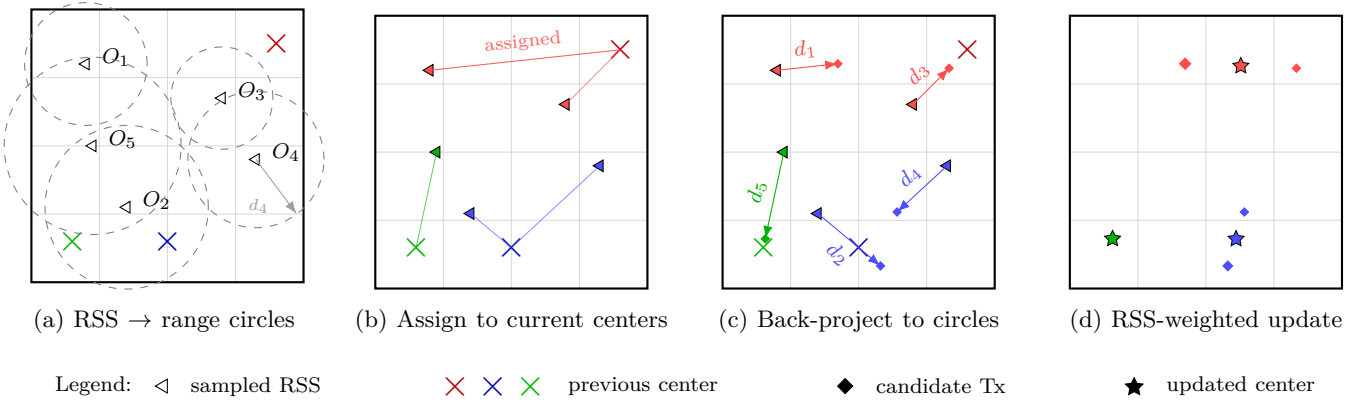
\begin{figure*}[t]
	\centering
	
	\begin{minipage}[b]{0.24\textwidth}
		\centering
		\begin{tikzpicture}[scale=0.9, every node/.style={font=\footnotesize}, >={Latex}]
			\draw[gray!30] (0,0) grid (4,4);
			\draw[thick] (0,0) rectangle (4,4);
			
			\coordinate (s1) at (0.8,3.2);
			\coordinate (s2) at (1.4,1.1);
			\coordinate (s3) at (2.8,2.7);
			\coordinate (s4) at (3.3,1.8);
			\coordinate (s5) at (0.9,2.0);
			
			\foreach \P/\name in {s1/$O_1$,s2/$O_2$,s3/$O_3$,s4/$O_4$,s5/$O_5$}{
				\node[regular polygon, regular polygon sides=3, draw, fill=white,
				minimum size=5pt, inner sep=0pt, rotate=90] at (\P) {};
				\node[anchor=west] at ($(\P)+(0.10,0.10)$) {\name};
			}
			
			\node[text=red!70!black]   at (3.6,3.5) {\Large $\times$};
			\node[text=blue!70!black]  at (2.0,0.6) {\Large $\times$};
			\node[text=green!70!black] at (0.6,0.6) {\Large $\times$};
			
			\draw[dashed,gray] (s1) circle (0.90);
			\draw[dashed,gray] (s2) circle (1.20);
			\draw[dashed,gray] (s3) circle (0.75);
			\draw[dashed,gray] (s4) circle (1.00);
			\draw[dashed,gray] (s5) circle (1.30);
			
			\coordinate (q4) at ($(s4)+(0.6,-0.8)$); 
			\draw[->,gray!75,line width=0.4pt] (s4) -- (q4);
			\node[gray!75,anchor=north east] at ($(s4)!0.55!(q4)$) {$\scriptstyle d_4$};
		\end{tikzpicture}
		
		\vspace{2pt}\centerline{\small (a) RSS $\rightarrow$ range circles}
	\end{minipage}
	\hfill
	\begin{minipage}[b]{0.24\textwidth}
		\centering
		\begin{tikzpicture}[scale=0.9, every node/.style={font=\footnotesize}, >={Latex}]
			\draw[gray!30] (0,0) grid (4,4);
			\draw[thick] (0,0) rectangle (4,4);
			
			\coordinate (cR) at (3.6,3.5);
			\coordinate (cB) at (2.0,0.6);
			\coordinate (cG) at (0.6,0.6);
			\node[text=red!70!black]   at (cR) {\Large $\times$};
			\node[text=blue!70!black]  at (cB) {\Large $\times$};
			\node[text=green!70!black] at (cG) {\Large $\times$};
			
			\coordinate (s1) at (0.8,3.2);
			\coordinate (s2) at (1.4,1.1);
			\coordinate (s3) at (2.8,2.7);
			\coordinate (s4) at (3.3,1.8);
			\coordinate (s5) at (0.9,2.0);
			
			\node[regular polygon, regular polygon sides=3, draw, fill=red!70,
			minimum size=5pt, inner sep=0pt, rotate=90] at (s1) {};
			\draw[red!70,opacity=.9] (s1)--(cR)
			node[midway,sloped,above,text=red!70] {\scriptsize assigned};
			
			\node[regular polygon, regular polygon sides=3, draw, fill=blue!70,
			minimum size=5pt, inner sep=0pt, rotate=90] at (s2) {};
			\draw[blue!70,opacity=.6] (s2)--(cB);
			
			\node[regular polygon, regular polygon sides=3, draw, fill=red!70,
			minimum size=5pt, inner sep=0pt, rotate=90] at (s3) {};
			\draw[red!70,opacity=.6] (s3)--(cR);
			
			\node[regular polygon, regular polygon sides=3, draw, fill=blue!70,
			minimum size=5pt, inner sep=0pt, rotate=90] at (s4) {};
			\draw[blue!70,opacity=.6] (s4)--(cB);
			
			\node[regular polygon, regular polygon sides=3, draw, fill=green!70!black,
			minimum size=5pt, inner sep=0pt, rotate=90] at (s5) {};
			\draw[green!70!black,opacity=.6] (s5)--(cG);
		\end{tikzpicture}
		
		\vspace{2pt}\centerline{\small (b) Assign to current centers}
	\end{minipage}
	\hfill
	\begin{minipage}[b]{0.24\textwidth}
		\centering
		\begin{tikzpicture}[scale=0.9, every node/.style={font=\footnotesize}, >={Latex}]
			\draw[gray!30] (0,0) grid (4,4);
			\draw[thick] (0,0) rectangle (4,4);
			
			\coordinate (cR) at (3.6,3.5);
			\coordinate (cB) at (2.0,0.6);
			\coordinate (cG) at (0.6,0.6);
			\node[text=red!70!black]   at (cR) {\Large $\times$};
			\node[text=blue!70!black]  at (cB) {\Large $\times$};
			\node[text=green!70!black] at (cG) {\Large $\times$};
			
			\coordinate (s1) at (0.8,3.2);
			\coordinate (s2) at (1.4,1.1);
			\coordinate (s3) at (2.8,2.7);
			\coordinate (s4) at (3.3,1.8);
			\coordinate (s5) at (0.9,2.0);
			
			\node[regular polygon, regular polygon sides=3, draw, fill=red!70,
			minimum size=5pt, inner sep=0pt, rotate=90] at (s1) {};
			\node[regular polygon, regular polygon sides=3, draw, fill=blue!70,
			minimum size=5pt, inner sep=0pt, rotate=90] at (s2) {};
			\node[regular polygon, regular polygon sides=3, draw, fill=red!70,
			minimum size=5pt, inner sep=0pt, rotate=90] at (s3) {};
			\node[regular polygon, regular polygon sides=3, draw, fill=blue!70,
			minimum size=5pt, inner sep=0pt, rotate=90] at (s4) {};
			\node[regular polygon, regular polygon sides=3, draw, fill=green!70!black,
			minimum size=5pt, inner sep=0pt, rotate=90] at (s5) {};
			
			\coordinate (p1) at ($(s1)!0.3196!(cR)$);
			\coordinate (p2) at ($(s2)!1.5364!(cB)$);
			\coordinate (p3) at ($(s3)!0.6629!(cR)$);
			\coordinate (p4) at ($(s4)!0.5652!(cB)$);
			\coordinate (p5) at ($(s5)!0.9080!(cG)$);
			
			\draw[->,red!70]          (s1) -- (p1) node[midway,sloped,above,text=red!70]   {$\scriptsize d_1$};
			\draw[->,blue!70]         (s2) -- (p2) node[midway,sloped,below,text=blue!70]  {$\scriptsize d_2$};
			\draw[->,red!70]          (s3) -- (p3) node[midway,sloped,above,text=red!70]   {$\scriptsize d_3$};
			\draw[->,blue!70]         (s4) -- (p4) node[midway,sloped,above,text=blue!70]  {$\scriptsize d_4$};
			\draw[->,green!70!black]  (s5) -- (p5) node[midway,sloped,above,text=green!70!black] {$\scriptsize d_5$};
			
			\foreach \P/\col in {p1/red!70,p2/blue!70,p3/red!70,p4/blue!70,p5/green!70!black}{
				\node[draw,\col,fill=\col,minimum size=2.2pt,inner sep=0pt,rotate=45] at (\P) {};
			}
		\end{tikzpicture}
		
		\vspace{2pt}\centerline{\small (c) Back-project to circles}
	\end{minipage}
	\hfill
	\begin{minipage}[b]{0.24\textwidth}
		\centering
		\begin{tikzpicture}[scale=0.9, every node/.style={font=\footnotesize}, >={Latex}]
			\draw[gray!30] (0,0) grid (4,4);
			\draw[thick] (0,0) rectangle (4,4);
			
			\node[draw,red!70,fill=red!70,minimum size=3.0pt,inner sep=0pt,rotate=45]  at (1.6949,3.2959) {};
			\node[draw,red!70,fill=red!70,minimum size=2.2pt,inner sep=0pt,rotate=45]  at (3.3303,3.2303) {};
			\node[draw,blue!70,fill=blue!70,minimum size=2.6pt,inner sep=0pt,rotate=45] at (2.3219,0.3318) {};
			\node[draw,blue!70,fill=blue!70,minimum size=2.2pt,inner sep=0pt,rotate=45] at (2.5652,1.1217) {};
			\node[draw,green!70!black,fill=green!70!black,minimum size=2.4pt,inner sep=0pt,rotate=45] at (0.6276,0.7289) {};
			
			\node[updstar, fill=red!70,  scale=7] at (2.5126,3.2631) {};
			\node[updstar, fill=blue!70, scale=7] at (2.4435,0.7267) {};
			\node[updstar, fill=green!70!black, scale=7] at (0.6276,0.7289) {};
			
		\end{tikzpicture}
		
		\vspace{2pt}\centerline{\small (d) RSS-weighted update}
	\end{minipage}
	
	\vspace{3mm}
	\begin{tikzpicture}[baseline, every node/.style={font=\footnotesize}]
		\matrix[matrix of nodes, row sep=0pt, column sep=2pt, nodes={anchor=west}]{
			\node{\textbf{Legend:}}; &
			\tikz \node[regular polygon, regular polygon sides=3, draw, fill=white, minimum size=6pt, inner sep=0pt, rotate=90] {}; & sampled RSS &
			\node[minimum width=12mm]{}; &
			{\color{red!70!black}\Large $\times$}\ {\color{blue!70!black}\Large $\times$}\ {\color{green!70!black}\Large $\times$} & previous center &
			\node[minimum width=12mm]{}; &
			\tikz \node[draw,fill=black,minimum size=4pt,inner sep=0pt,rotate=45] {}; & candidate Tx &
			\node[minimum width=12mm]{}; &
			\tikz \node[updstar, fill=black,  scale=7] {}; & updated center \\
		};
	\end{tikzpicture}
	
	\caption{Four-step visualization of the propagation-guided initialization. (a) Convert each sampled RSS to a distance circle. (b) Keep the receiver shape and color by assigned center. (c) Keep colored receivers and back-project to the circle. (d) Update centers by an RSS-weighted centroid. Colors indicate clusters.}
	\label{fig:pgkmeans_four_minipage_colored_clear2}
\end{figure*}

By the chain rule, $\frac{\partial \mathcal{L}_t}{\partial a_t^i}$ is computed as
\begin{equation}\label{eq:chain}
	\frac{\partial \mathcal{L}_t}{\partial a_t^i} = \frac{\partial \mathcal{L}_t}{\partial \hat{{\bm x}}_0^t} \cdot \frac{\partial \hat{\bm x}_0^t}{\partial {{\bm C}}_t}  \cdot \frac{\partial {{\bm C}}_t}{\partial a_t^i} + \kappa_t \left( a^i_t - a^{i,\star}_t \right),
\end{equation}
where $a^{i,\star}_t$ is the horizontal coordinate of $\Omega_t^\star$. 
{\black The $\frac{\partial \mathcal{L}_t}{\partial \hat{{\bm x}}_0^t}$ and $\frac{\partial \hat{\bm x}_0^t}{\partial {{\bm C}}_t}$ can be easily obtained by back-propagation}.
As for $\frac{\partial {{\bm C}}_t}{\partial a_t^i}$, note that the binary Tx map ${\bm C}_t$ is obtained by applying a non-differentiable operation, cf. Eqn.~\eqref{eq:hardC}. This makes the gradient ${\partial {{\bm C}}_t}/{\partial a_t^i}$ undefined.
To circumvent this difficulty, we introduce a smooth approximation $\hat{{\bm C}}_t \in \mathbb{R}_+^{w \times h}$, formed by summing $R$ isotropic Gaussian heatmaps centered at each Tx location:
\begin{equation}\label{eq:gaussian}
	\hat{{\bm C}}_t = \sum_{i=1}^{R} {\bm P}_t^i, \quad {\bm P}_t^i(j, k) = \exp\left( -\frac{(j - a_t^i)^2 + (k - b_t^i)^2}{2\sigma^2} \right),
\end{equation}
where $\sigma$ is the standard deviation, {\black which controls approximation accuracy. Particularly as $\sigma$ approaches 0, the soft map $\hat{{\bm C}}_t$ converges to the binary Tx map ${\bm C}_t$.}

With this approximation, we adopt the straight-through estimator (STE)~\cite{bengio2013estimating, NIPS2017_7a98af17} and approximate the backward gradient by differentiating through the soft map:
\begin{equation}
	\frac{\partial {{\bm C}}_t}{\partial a_t^i} \approx \frac{\partial \hat{{\bm C}}_t}{\partial a_t^i}
	= \frac{\partial {\bm P}_t^i}{\partial a_t^i}
	= \frac{1}{\sigma^2} ({\bm J - a_t^i}) \odot {\bm P}_t^i,
\end{equation}
where $\bm J \in \mathbb{R}^{w \times h}$ is a row index matrix satisfying ${\bm J}(j,k) = j$.

Thus far, we have completed the description of the two core components—RM generation and Tx update. Since the RM estimation problem is highly non-convex, the overall procedure is notably sensitive to the initialization of the Tx coordinates. In the following subsection, we introduce an efficient strategy to obtain a reliable initialization for $\Omega_T$.

\subsection{Initialization for Tx Coordinates $\Omega_T$ }\label{sec:pgkmeans}
We propose a propagation-guided k-means (PG-KMeans) method to initialize the coordinates at $t=T$. The one-iteration visualization of the method is shown in Fig.~\ref{fig:pgkmeans_four_minipage_colored_clear2}. Let $\mathcal{S}=\{(m_k,n_k)\}$ be the set of sampled grid indices with $\mathcal{M}(m_k,n_k){=}1$, and recall that ${\bm O}(m_k,n_k)$ denote the measured RSS at those indices. Using a site-calibrated first-order path-loss model $o=p(d;\vartheta)$ with known parameters $\vartheta$, which is easy to get by preliminary investigation, we can map each RSS value to a range via the analytic inverse $d=\psi(o)=p^{-1}(o;\vartheta)$:
\begin{equation}
	d_k=\psi\big({\bm O}(s_k)\big),\qquad {\bf s}_k=\big(m_k,n_k\big) ,
\end{equation}
so that each sample $k$ specifies a circle of feasible Tx locations centered at ${\bf s}_k$ with radius $d_k$. Stacking the $R$ Tx positions into $\Omega=\{\boldsymbol\omega_i\in\mathbb{R}^2\}_{i=1}^{R}$, we seek $\Omega$ that best fits all circles in a least-squares sense. A convenient surrogate is the range-consistency distortion
\begin{equation}
	\mathcal{F}(\Omega)\;=\;\sum_{k\in\mathcal{S}} \min_{1\le i\le R}\Big(\,\big\|{\bf s}_k-\boldsymbol\omega_i\big\|_2 - d_k\Big)^2 ,
	\label{eq:init_range_obj_text}
\end{equation}
which measures how well each sample can be explained by its nearest Tx. Minimizing \eqref{eq:init_range_obj_text} is nonconvex but admits a fast alternating strategy that mirrors K-means while respecting propagation geometry.

\begin{algorithm}[t]
	\caption{RadioTrace\label{alg:radiotrace}}
	\textbf{Input:} Sparse RSS $\bm O$, mask $\mathcal M$, building map $\bm B$, \#Tx $R$, diffusion model $f_\theta$, steps $T$, step size $\eta$, momentum $\beta$, anchor cap $\kappa$, Gaussian width $\sigma$, PG-KMeans iters $L_{\max}$, tol $\tau$ \\
	\textbf{Output:} Clean radio map $\bm x_0$, Tx coordinates $\Omega_0$
	\begin{algorithmic}[1]
		\State \textbf{PG-KMeans init.:} $\mathcal S \gets \{(m_k,n_k)\,|\,\mathcal M{=}1\}$
		\State Map RSS to ranges: $d_k \gets \psi\!\big(\bm O(m_k,n_k)\big)$
		\State Init seeds $\{\boldsymbol\omega_i^{(0)}\}_{i=1}^{R}$ uniformly on grid
		\For{$\ell = 0$ to $L_{\max}-1$}
		\State Assign $k$ to $i = \arg\min_i \big|\|s_k{-}\boldsymbol\omega_i^{(\ell)}\|{-}d_k\big|$
		\State Back-project: $\tilde{\boldsymbol\omega}_{k,i} \gets s_k + d_k\frac{\boldsymbol\omega_i^{(\ell)}-s_k}{\max\{\|\boldsymbol\omega_i^{(\ell)}-s_k\|,\varepsilon\}}$
		\State Update: $\boldsymbol\omega_i^{(\ell+1)} \gets \frac{\sum_{k\in \mathcal{E}_i}\tilde O_k \tilde{\boldsymbol\omega}_{k,i}}{\sum_{k\in \mathcal{E}_i}\tilde O_k}$
		\State \textbf{if} $\max_i\|\boldsymbol\omega_i^{(\ell+1)}{-}\boldsymbol\omega_i^{(\ell)}\|\le\tau$ \textbf{then break}
		\EndFor
		\State $\Omega_T \gets \{\boldsymbol\omega_i^{(\ell_{\text{end}})}\}_{i=1}^{R}$; sample $\bm x_T \sim \mathcal N(\bm 0,\bm I)$
		\State Set anchor $\Omega_T^\star \gets \Omega_T$; best loss $\mathcal L^\star \gets +\infty$
		\State Initialize momentum $v_{a,i}{=}v_{b,i}{=}0$ for $i{=}1{:}R$
		
		\For{$t = T$ \textbf{to} $1$}
		\State Build binary Tx map ${\bm C}_t$ with $\Omega_t$
		\State Predict noise: $\epsilon_\theta \gets f_\theta(\bm x_t,t,\bm B,{\bm C}_t)$
		\State $\hat{\bm x}_0^t \gets \frac{1}{\sqrt{\bar\alpha_t}}\!\left(\bm x_t - \sqrt{1{-}\bar\alpha_t}\,\epsilon_\theta\right)$
		\State Sample $\bm x_{t-1}$ with DDPM mean and variance
		\State $\kappa_t \gets \kappa \cdot \frac{T-t}{T}$
		\State Loss: $\mathcal{L}_t \gets \|(\hat{\bm x}_0^t \odot \mathcal M)-\bm O\|_F^2 + \frac{\kappa_t}{2}\,\big\|\Omega_t-\Omega_t^\star\big\|_2^2$
		\State Backprop to get $\nabla_{\Omega_t}\mathcal{L}_t$
		\State Update the coordinates to get $\Omega_{t-1}$
		\State \textbf{if} $\mathcal{L}_t < \mathcal{L}^\star$ \textbf{then} $\mathcal L^\star\!\gets\!\mathcal{L}_t$, $\Omega_{t-1}^\star\!\gets\!\Omega_{t}$
		\EndFor
		
		\State \Return $\bm x_0$, $\Omega_0$
	\end{algorithmic}
\end{algorithm}

Starting from random seeds $\{\boldsymbol\omega_i^{(0)}\}_{i=1}^{R}$, we iterate three simple operations:
\begin{enumerate}
	\item \textit{Assignment:} For each sample $k$, compute
	$e_{k,i}=\big|\|{\bf s}_k-\boldsymbol\omega_i^{(\ell)}\|_2-d_k\big|$
	and assign $k$ to cluster $\mathcal{E}_i^{(\ell)}$ that attains the minimum $e_{k,i}$.
	\item \textit{Back-projection:} For each $(k,i)$ with $k\in \mathcal{E}_i^{(\ell)}$, form a candidate Tx point by projecting ${\bf s}_k$ onto its range circle along the ray toward the current center:
	\begin{equation}
		\tilde{\boldsymbol\omega}_{k,i}
		\,=\, {\bf s}_k \;+\; d_k\cdot 
		\frac{\boldsymbol\omega_i^{(\ell)}-{\bf s}_k}{\max\{\|\,\boldsymbol\omega_i^{(\ell)}-{\bf s}_k\,\|_2,\ \epsilon\}},
		\label{eq:init_backproj}
	\end{equation}
	with a small $\epsilon>0$ to avoid division by zero.
	\item \textit{Update:} use a confidence-weighted centroid with weights taken directly from RSS,
	\begin{equation}
		\boldsymbol\omega_i^{(\ell+1)}
		\;=\;
		\frac{\sum_{k\in \mathcal{E}_i^{(\ell)}} \widetilde{O}_k\,\tilde{\boldsymbol\omega}_{k,i}}
		{\sum_{k\in \mathcal{E}_i^{(\ell)}} \widetilde{O}_k},
		\label{eq:init_update_rss}
	\end{equation}
	where $\epsilon>0$ is a small constant and $\widetilde{O}_k$ is a non-negative proxy of the sampled RSS at ${\bf s}_k$:
	\[
	\widetilde{O}_k \;=\; \log ({1 + \exp{({\bm O}(m_k,n_k))}}).
	\]
\end{enumerate}
We stop when $\max_i\|\boldsymbol\omega_i^{(\ell+1)}-\boldsymbol\omega_i^{(\ell)}\|_2\le \tau$ or when $\ell$ matches the max iterations $L_{\max}$. This propagation-guided K-means decreases \eqref{eq:init_range_obj_text} in practice and typically converges within $10$ iterations by our numerical experience. Larger RSS indicates stronger links and typically shorter ranges, so weighting by $\widetilde{O}_k$ prioritizes more informative samples while keeping the mechanics simple.

Putting all pieces together, the complete RadioTrace procedure is summarized in Algorithm~\ref{alg:radiotrace}.

{\black 
\section{Theoretical Analysis} \label{sec:theory}

{\black This section analyzes the stability of the Tx-coordinate refinement component in the proposed RadioTrace framework. Due to the tightly coupled in-loop inference procedure, the effective refinement objective evolves across diffusion time steps, since the RM estimate and the relaxed Tx-map representation are progressively updated during reverse generation. We therefore formulate the Tx-coordinate update as a time-varying stochastic optimization process and analyze its long-term behavior under standard regularity conditions.}

For the purpose of convergence analysis, we study the algorithm in the forward direction with time index $k=0,1,2,\ldots$ instead of the diffusion reverse iteration, and rewrite the coordinate-refinement recursion accordingly. This re-indexing does not alter the algorithm but simplifies the subsequent analysis.
At analysis step $k$, the reverse diffusion process produces a random RM reconstruction, which induces a stochastic loss function
$
L_k(\Omega;\xi_k),
$
where $\Omega\in\mathbb{R}^{2R}$ denotes the Tx-coordinate vector and $\xi_k$ is a random variable, arising from the diffusion sampling and the rounding procedures. Let
$
\mathcal{F}_k:=\mathcal{F}(\Omega_0,\xi_0,\ldots,\xi_{k-1})
$
be the natural filtration generated by the past iterates and past randomness, and define the conditional mean loss
\[
\bar L_k(\Omega):=\mathbb{E}\!\left[L_k(\Omega;\xi_k)\mid\mathcal{F}_k\right].
\]

The refinement update is rewritten as
\begin{equation} \label{eq:update}
	v_{k+1}=\beta v_k+(1-\beta)G_k,\qquad
	\Omega_{k+1}=\Omega_k-\eta_k v_{k+1},
\end{equation}
where $G_k$ denotes the implemented gradient surrogate. 
We next introduce the assumptions required for the subsequent analysis.

\begin{assumption}[Smoothness]\label{ass:smooth}
For each $k$, the function $\bar L_k$ is $M_k$-smooth, namely,
\[
\|\nabla \bar L_k(\Omega_1)-\nabla \bar L_k(\Omega_2)\|
\le M_k\|\Omega_1-\Omega_2\|,\qquad \forall \Omega_1,\Omega_2.
\]
\end{assumption}

\begin{assumption}[Bounded Drift and Perturbation]\label{ass:drift}
The implemented gradient admits the decomposition
\[
G_k=\nabla \bar L_k(\Omega_k)+\zeta_k,
\qquad
m_k:=\mathbb{E}[\zeta_k\mid\mathcal{F}_k].
\]
where $\zeta_k$ models the combined effect of diffusion sampling and rounding on the gradient. 
There exist non-negative sequences $\{\delta_k\}$, $\{\sigma_k\}$, and $\{\bar d_k\}$, and a constant $\bar\delta>0$, such that for all $k$,
\[
\|m_k\|\le \delta_k\le \bar\delta,
\qquad
\mathbb{E}\!\left[\|\zeta_k-m_k\|^2\mid\mathcal{F}_k\right]\le \sigma_k^2,
\]
\[
\big|\bar L_{k+1}(\Omega)-\bar L_k(\Omega)\big|\le \bar d_k, ~ \forall \Omega,
\]
and
\[
\sum_{k=0}^{\infty}\bar d_k<\infty,
\qquad
\sum_{k=0}^{\infty}\eta_k\delta_k<\infty,
\qquad
\sum_{k=0}^{\infty}\eta_k\sigma_k^2<\infty.
\]
\end{assumption}

\begin{assumption}[Stepsizes and momentum]\label{ass:steps}
The momentum parameter satisfies $\beta\in[0,1)$,
and the stepsizes $\{\eta_k\}$ are nonincreasing such that
\[
\sum_{k=0}^{\infty}\eta_k=\infty,\qquad
\eta_k\to 0,\qquad
\eta_kM_k\to 0.
\]
\end{assumption}
{\black The above assumptions should be interpreted as regularity conditions for the Tx-coordinate refinement process under reverse diffusion. Assumptions~\ref{ass:smooth} and~\ref{ass:steps} are standard technical conditions in stochastic approximation: the former imposes smoothness on the conditional mean refinement loss over a finite spatial search region, while the latter specifies a classical diminishing stepsize rule. The key condition is Assumption~\ref{ass:drift}. It requires the accumulated objective drift, gradient perturbation, and stochastic fluctuation caused by reverse diffusion to remain finite.
This condition is reasonably aligned with the practical denoising process, in which the noise scale $\gamma_t$ gradually decreases and the randomness of the refinement objective is expected to diminish over time. The empirical behavior of the Tx-coordinate refinement process reported in Section V-L also provides supporting evidence for this controlled-drift interpretation.}

We are now ready to state the main theorem, which summarizes the long-term behavior of the algorithm under the above conditions.

\begin{theorem}\label{thm:main}
Under Assumptions 1--3, the iterates, generated by \eqref{eq:update}, satisfy
\[
\liminf_{k\to\infty}\mathbb{E}\!\left[\|\nabla \bar L_k(\Omega_k)\|^2\right]=0.
\]
\end{theorem}

The proof is based on a Lyapunov argument. By exploiting the smoothness of $\bar L_k$
and the momentum recursion, a one-step descent inequality can be established for a suitable Lyapunov function, which together with bounded drift assumption gives the stated convergence result. The detailed proof is provided in Appendix~\ref{app:thm}.

{\black Theorem~\ref{thm:main} establishes that the expected squared gradient of the conditional mean loss asymptotically approaches zero, demonstrating the stability of Tx-coordinate refinement under time-varying objectives and stochastic perturbations.
}

}

\section{Experiments} \label{sec:Exp}
\subsection{Dataset}
In this study, we evaluate the performance of the proposed method using the RadioMapSeer~\cite{Levie2021RadioUNet} {\black and the BART-Lab dataset~\cite{zhang2024physics}}. 

The \textbf{RadioMapSeer} dataset contains 700 building maps, each with 80 Tx locations. These building maps cover major cities such as Ankara, Berlin, and so on. The heights of Tx, receivers, and buildings are defined as 1.5 meters (m), 1.5 m, and 25 m, respectively. The Tx power is set to 23 dBm, and the carrier frequency is 5.9 GHz. The RM are generated using Maxwell's equations and stored as 256$\times$256 m$^2$ grids with a 1-meter resolution. 
Among the 700 building maps, we use 500 maps for training the pre-trained measurement-free RM generation model used in our method and the deep-learning-based RM estimation baselines, while the remaining 200 maps are reserved for evaluating all methods.
Because RadioMapSeer provides only single-Tx RMs (in dBm), we synthesize multi-Tx scenes for evaluation by converting each single-Tx map to linear power, summing the maps pixel-wise across multiple distinct Tx selections from the same building, converting the composite back to dBm. To ensure a controlled comparison and isolate geometric/propagation effects, we set all Tx to the same power (23 dBm) when composing multi-Tx scenes; thus, variation arises from layout and occlusions rather than per-Tx power differences.

{\black The \textbf{BART-Lab} is a more challenging dataset than RadioMapSeer, due to its finer-grained building layouts and more intricate propagation characteristics. The dataset is generated using Altair Feko over regions in the United States, where the environmental maps are obtained from OpenStreetMap. In each scenario, buildings are modeled with a height of 10 m, and three Tx are deployed, each with antennas mounted at 30 m and an initial transmit power of 46.00 dBm. Among the five available frequency bands, we use the 1750 MHz band in our experiments for consistency. Furthermore, from each original radio map, we extract a 256$\times$256 subregion that covers all three Tx, thereby preserving the essential multi-Tx propagation characteristics.}

\subsection[Setup]{Setup\footnote{\black The source code is available at \href{https://github.com/YannLeo/RadioTrace}{https://github.com/YannLeo/RadioTrace}}}

\begin{figure}
	%
	\begin{minipage}[b]{0.48\linewidth}
		\centering
		\centerline{\includegraphics[width=3.8cm]{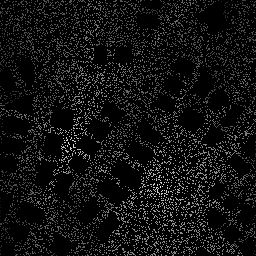}}
		\centerline{(a) Random Sampling}\medskip
	\end{minipage}
	\hfill
	\begin{minipage}[b]{0.48\linewidth}
		\centering
		\centerline{\includegraphics[width=3.8cm]{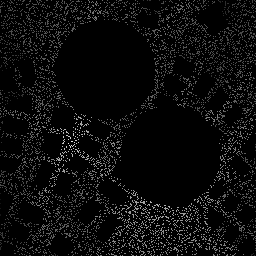}}
		\centerline{(b) Restricted-area Sampling}\medskip
	\end{minipage}
	\caption{The sampled RSS from different sampling mode with sampling rate of 20\%.}
	\label{fig:sampling}     
\end{figure}

\begin{figure*}[htb]
	
	\centering
	\centerline{\includegraphics[width=18cm]{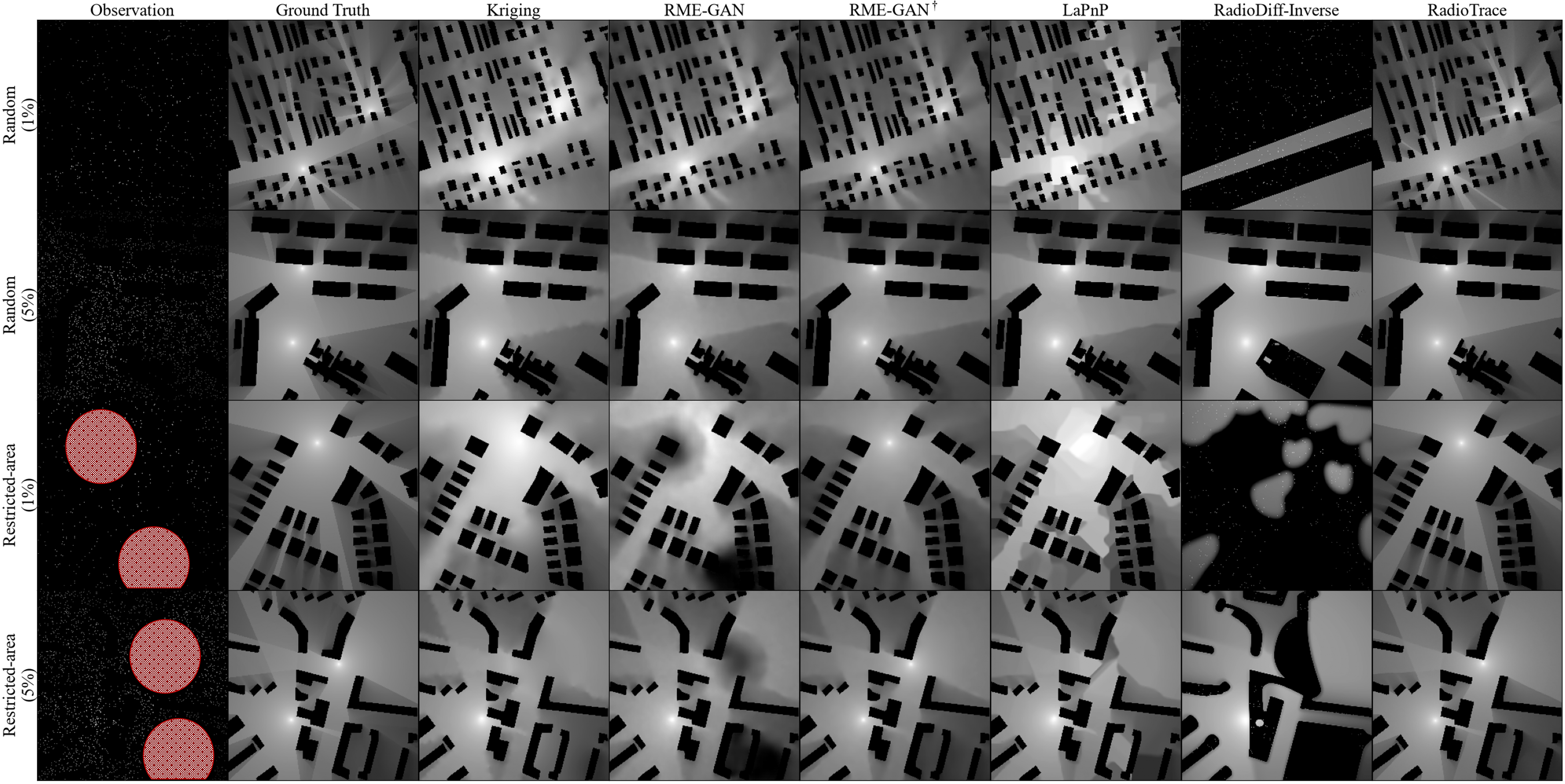}}
	
	\caption{\black Comparison of RM Estimation on RadioMapSeer dataset under \textbf{random sampling} (row 1, 2) and \textbf{restricted-area sampling} (row 3, 4) with low sampling ratios (1\% and 5\%). The red regions correspond to restricted areas where no measurements are available.}
	\label{fig:results}
\end{figure*}

\subsubsection{Sampling modes} \label{sec:setup}

To evaluate the performance of RadioTrace under different measurement conditions, we consider two sampling modes for collecting RSS measurements in outdoor environments:
\begin{itemize}
	\item \textbf{Random sampling}: RSS measurements are uniformly and randomly collected from the entire area without any spatial constraints.
	\item \textbf{Restricted-area sampling}: It defines some restricted areas, within which no RSS measurements can be collected. 
	Outside the restricted regions, RSS measurements are uniformly and randomly sampled across the accessible area.
	Restricted-area sampling is much more challenging than random sampling case, especially when there exists Tx inside the restricted areas.
	
\end{itemize}

\subsubsection{Implementation Details}
For the diffusion-based measurement-free RM generation model used in RadioTrace, we adopt the pre-trained model provided by RadioDiff~\cite{Wang2024RadioDiff}, without any additional fine-tuning or retraining. The RadioDiff backbone comprises a U-Net core~\cite{ronneberger2015u} with approximately 137M parameters and a Swin-Transformer–based conditional embedding network~\cite{liu2021swin} with about 87M parameters. 
The reverse denoising process is performed with $T=100$ steps.
For the Tx map generation, we set the standard deviation of the Gaussian kernel to $\sigma=10$. The Tx coordinates are optimized using momentum-based GD, with a learning rate of $\eta=300$ and a momentum coefficient of $\beta=0.4$. The upper bound of the linear scheduler of incumbent best anchor $\kappa=0.8$. The max iterations of PG-KMeans $L_{\max}=10$. The Tx coordinate set before PG-KMeans is initialized by selecting the $R$ locations with the highest RSS values.

\subsubsection{Baselines}
\label{subsubsec:baselines_assumptions}

To better contextualize the reported gains, we briefly introduce the main baselines used in our experiments.
\begin{itemize}
	\item \textbf{Kriging}~\cite{Romero2022Radio} formulates RM estimation as spatial interpolation from sparse RSS samples via a hand-crafted covariance model. It does not use Tx locations or learned generative priors.
	\item \textbf{RME-GAN}~\cite{Zhang2023RMEGAN} is a supervised, data-driven model trained offline to learn a feed-forward mapping from sparse measurements to full radio maps. It relies on large-scale training data and assumes Tx locations are available as part of the conditioning setup in our comparison. {\black By default, RME-GAN is trained under the random sampling setting. In addition, we consider a stronger variant trained under both random and restricted-area sampling settings, denoted as RME-GAN$^\dagger$, to evaluate the effect of sampling-pattern-specific supervision.}
	\item \textbf{LaPnP}~\cite{Xu2025Radio} treats RM estimation as an inverse problem by alternating between enforcing measurement consistency and applying a pre-trained latent-domain prior. Tx locations are not utilized during inference, and the prior only serves as a regularizer for map completion.
	{\black \item \textbf{RadioDiff-Inverse}~\cite{wang2025radiodiff} treats RM construction as a diffusion-enhanced Bayesian inverse problem from noisy sparse measurements. Tx locations are not explicitly utilized during inference, and the learned diffusion prior mainly serves as a distributional regularizer rather than a Tx-geometry-aware reconstruction guide.}

\end{itemize}

\subsection{Metrics}
We evaluate reconstruction quality with four metrics: normalized mean-square error (NMSE), root-mean-square error (RMSE), structural similarity (SSIM), and peak signal-to-noise ratio (PSNR). For each test scene, we first linearly rescale the ground-truth $\bm{x}$ and predicted dB radio maps $\hat{\bm{x}}$ to $[0,1]$, then compute all metrics on the rescaled maps.

NMSE measures the squared error relative to the energy of the ground-truth map, providing a scale-invariant notion of distortion across scenes
\begin{equation}
	\mathrm{NMSE} \;=\; \frac{\lVert \hat{\bm{x}}-\bm{x}\rVert_2^2}{\lVert \bm{x}\rVert_2^2}.
\end{equation}

 RMSE is the average per-pixel deviation in the same $[0,1]$ units as the rescaled maps and is easy to interpret as an absolute error level
\begin{equation}
	\mathrm{RMSE} \;=\; \sqrt{\frac{1}{N}\sum_{i=1}^{N}\bigl(\hat{x}_i-x_i\bigr)^2}.
\end{equation}

SSIM focuses on structural fidelity by comparing local luminance, contrast, and structure; values typically lie in $[0,1]$ with higher being better
\begin{equation}
	\mathrm{SSIM}(\bm{x},\hat{\bm{x}})
	\;=\;
	\frac{\bigl(2\,\mu_{\bm{x}}\,\mu_{\hat{\bm{x}}}+c_1\bigr)\bigl(2\,\sigma_{\bm{x}\hat{\bm{x}}}+c_2\bigr)}
	{\bigl(\mu_{\bm{x}}^{2}+\mu_{\hat{\bm{x}}}^{2}+c_1\bigr)\bigl(\sigma_{\bm{x}}^{2}+\sigma_{\hat{\bm{x}}}^{2}+c_2\bigr)} .
\end{equation}
Here $\mu_{\bm{x}}$ and $\mu_{\hat{\bm{x}}}$ are local means, $\sigma_{\bm{x}}^{2}$ and $\sigma_{\hat{\bm{x}}}^{2}$ are local variances, $\sigma_{\bm{x}\hat{\bm{x}}}$ is the local covariance, and $c_1, c_2$ are stabilization constants.

PSNR expresses fidelity in decibels as a monotonic transform of MSE; with maps rescaled to $[0,1]$, larger PSNR directly corresponds to smaller MSE
\begin{equation}
	\mathrm{PSNR}
	\;=\; 10\log_{10}\!\left(\frac{1}{\mathrm{MSE}}\right).
\end{equation}
The metrics are computed over all pixels per scene and then averaged over the test set.


\subsection{Results of comparable experiments}

\subsubsection{\black Results on RadioMapSeer dataset}
We first evaluate the performance of different methods on RadioMapSeer dataset under the random sampling pattern introduced in Section~\ref{sec:setup}. We compare our proposed RadioTrace with several representative RM estimation baselines mentioned in Section~\ref{subsubsec:baselines_assumptions}.

In this experiment, the RSS measurements are uniformly and randomly collected from the entire area without any spatial constraints. We evaluate all methods under different sampling ratios, which denote the proportion of observed locations. Specifically, we consider four sampling ratios: 1\%, 5\%, 10\%, and 20\%, to simulate various levels of observation sparsity.

The quantitative comparison under random sampling is summarized in Table~\ref{tab:random}. RadioTrace clearly surpasses the non–learning baselines (Kriging and LaPnP) at all sampling rates, delivering higher perceptual fidelity and lower distortion. {\black The results of RadioDiff-Inverse indicate that posterior reconstruction with an unconditional diffusion prior may have difficulty capturing Tx-dependent propagation structures under sparse RSS observations.} Compared with RME-GAN, RadioTrace attains the best RMSE, SSIM, and PSNR across all rates while remaining second-best in NMSE. For example, at 1\% sampling RadioTrace improves PSNR from 30.98\,dB to 32.92\,dB and raises SSIM from 0.9457 to 0.9590; at 20\% it further widens the PSNR gap (32.79\,dB vs.\ 33.71\,dB) with higher SSIM (0.9613 vs.\ 0.9645). 

The reconstructed RMs under random sampling with low sampling ratios (1\% and 5\%) are visualized in Row 1 and Row 2 of Fig.~\ref{fig:results}. As shown, RadioTrace produces reconstructions that are visually closer to the ground truth, with clearer signal propagation boundaries and fewer artifacts compared to Kriging and LaPnP. Although RME-GAN also yields good results under random sampling due to supervised training, its performance is slightly inferior in terms of structure preservation. {\black RadioDiff-Inverse fails to recover a meaningful RM at the 1\% sampling rate. At 5\%, the relatively large spacing between sparse observations limits its ability to resolve localized propagation variations, leading to inaccurate reconstruction near building boundaries. Overall, RadioTrace better preserves both the global propagation pattern and local structural details.}

We further evaluate the performance of different methods under the restricted-area sampling mode introduced in Section~\ref{sec:setup}, where RSS measurements are unavailable in specific regions due to environmental constraints. In this experiment, we define two circular restricted areas, each with a radius of 50 meters, within which no measurements can be collected. The remaining accessible region is used for RSS sampling, where measurements are uniformly and randomly drawn. 
We consider four sampling ratios: 1\%, 5\%, 10\%, and 20\%.


\begin{table}[t] 
	\scriptsize
	\belowrulesep=0pt
	\aboverulesep=0pt
	\renewcommand\arraystretch{1.1}
	\setlength{\tabcolsep}{3pt}
	\centering
	\caption{\label{tab:random}\black Results of random sampling on RadioMapSeer}
	\resizebox{\linewidth}{!}{
		\begin{tabular}{c | c | c | c | c | c | c}
			\toprule
			\textbf{Sampling rate} & \textbf{Metrics} & Kriging & RME-GAN & LaPnP & {\black RadioDiff-Inverse} & RadioTrace \\
			\hline
			\multirow{4}{*}{1\%}   
			& NMSE             & 0.0274 & \textbf{0.0062}     & 0.0192 & {\black 0.5835}  & \underline{0.0093} \\
			& RMSE             & 0.0524 & \underline{0.0282}  & 0.0433 & {\black 0.2478}  & \textbf{0.0244} \\
			& SSIM $\uparrow$  & 0.8726 & \underline{0.9457}  & 0.9056 & {\black 0.3322}  & \textbf{0.9590} \\
			& PSNR $\uparrow$  & 25.81  & \underline{30.98}   & 27.48  & {\black 12.29} & \textbf{32.92} \\
			\hline
			\multirow{4}{*}{5\%}   
			& NMSE             & 0.0233 & \textbf{0.0046}     & 0.0138 & {\black 0.0841}  & \underline{0.0070} \\
			& RMSE             & 0.0481 & \underline{0.0243}  & 0.0364 & {\black 0.0911}  & \textbf{0.0229} \\
			& SSIM $\uparrow$  & 0.8929 & \underline{0.9573}  & 0.9358 & {\black 0.7588}  & \textbf{0.9616} \\
			& PSNR $\uparrow$  & 26.58  & \underline{32.28}   & 29.03  & {\black 21.09} & \textbf{33.26} \\
			\hline
			\multirow{4}{*}{10\%}  
			& NMSE             & 0.0226 & \textbf{0.0043}     & 0.0130 & {\black 0.0586}  & \underline{0.0065} \\
			& RMSE             & 0.0473 & \underline{0.0235}  & 0.0353 & {\black 0.0597}  & \textbf{0.0224} \\
			& SSIM $\uparrow$  & 0.8996 & \underline{0.9600}  & 0.9423 & {\black 0.8724}  & \textbf{0.9630} \\
			& PSNR $\uparrow$  & 26.73  & \underline{32.56}   & 29.31  & {\black 24.88} & \textbf{33.44} \\
			\hline
			\multirow{4}{*}{20\%}  
			& NMSE             & 0.0221 & \textbf{0.0041}     & 0.0123 & {\black 0.0188}  & \underline{0.0054} \\
			& RMSE             & 0.0469 & \underline{0.0229}  & 0.0344 & {\black 0.0435}  & \textbf{0.0215} \\
			& SSIM $\uparrow$  & 0.9048 & \underline{0.9613}  & 0.9471 & {\black 0.9257}  & \textbf{0.9645} \\
			& PSNR $\uparrow$  & 26.81  & \underline{32.79}   & 29.53  & {\black 27.46} & \textbf{33.71} \\
			\bottomrule
		\end{tabular}
	}
\end{table}

\begin{table}[t] 
	\scriptsize
	\belowrulesep=0pt
	\aboverulesep=0pt
	\renewcommand\arraystretch{1.1}
	\setlength{\tabcolsep}{2.5pt}
	\centering
	\caption{\label{tab:restricted}\black Results of restricted-area sampling on RadioMapSeer}
	\resizebox{\linewidth}{!}{
		\begin{tabular}{c | c | c | c | c | c | c | c}
			\toprule
			\textbf{Sampling rate} & \textbf{Metrics} & Kriging & RME-GAN & {\black RME-GAN$^\dagger$} & LaPnP & {\black RadioDiff-Inverse} & RadioTrace \\
			\hline
			\multirow{4}{*}{1\%}   
			& NMSE             & 0.0392 & 0.0715 & {\black \textbf{0.0096}}    & 0.0336 & {\black 0.7108} & \underline{0.0174} \\
			& RMSE             & 0.0622 & 0.0958 & {\black \underline{0.0335}} & 0.0576 & {\black 0.2808} & \textbf{0.0296} \\
			& SSIM $\uparrow$  & 0.8643 & 0.8829 & {\black \underline{0.9301}} & 0.8860 & {\black 0.2790} & \textbf{0.9518} \\
			& PSNR $\uparrow$  & 24.32  & 20.37  & {\black \underline{29.55}}  & 24.99  & {\black 11.15}  & \textbf{31.95} \\
			\hline
			\multirow{4}{*}{5\%}   
			& NMSE             & 0.0331 & 0.0521 & {\black \textbf{0.0078}}    & 0.0267 & {\black 0.2697} & \underline{0.0141} \\
			& RMSE             & 0.0573 & 0.0818 & {\black \underline{0.0302}} & 0.0513 & {\black 0.1689} & \textbf{0.0266} \\
			& SSIM $\uparrow$  & 0.8818 & 0.9069 & {\black \underline{0.9421}} & 0.9139 & {\black 0.6490} & \textbf{0.9566} \\
			& PSNR $\uparrow$  & 25.05  & 21.74  & {\black \underline{30.43}}  & 25.99  & {\black 15.71}  & \textbf{32.57} \\
			\hline
			\multirow{4}{*}{10\%}  
			& NMSE             & 0.0320 & 0.0489 & {\black \textbf{0.0075}}    & 0.0255 & {\black 0.1944} & \underline{0.0115} \\
			& RMSE             & 0.0563 & 0.0792 & {\black \underline{0.0297}} & 0.0502 & {\black 0.1415} & \textbf{0.0255} \\
			& SSIM $\uparrow$  & 0.8878 & 0.9114 & {\black \underline{0.9451}} & 0.9218 & {\black 0.7609} & \textbf{0.9586} \\
			& PSNR $\uparrow$  & 25.19  & 22.02  & {\black \underline{30.57}}  & 26.19  & {\black 17.41}  & \textbf{32.81} \\
			\hline
			\multirow{4}{*}{20\%}  
			& NMSE             & 0.0314 & 0.0471 & {\black \textbf{0.0074}}    & 0.0249 & {\black 0.1622} & \underline{0.0108} \\
			& RMSE             & 0.0558 & 0.0778 & {\black \underline{0.0296}} & 0.0496 & {\black 0.1280} & \textbf{0.0245} \\
			& SSIM $\uparrow$  & 0.8922 & 0.9136 & {\black \underline{0.9460}} & 0.9244 & {\black 0.8176} & \textbf{0.9605} \\
			& PSNR $\uparrow$  & 25.29  & 22.18  & {\black \underline{30.61}}  & 26.30  & {\black 18.41}  & \textbf{33.09} \\
			\bottomrule
		\end{tabular}
	}
\end{table}

\begin{table}[t]
	\scriptsize \black
	\belowrulesep=0pt
	\aboverulesep=0pt
	\renewcommand\arraystretch{1.1}
	\setlength{\tabcolsep}{2.5pt}
	\centering
	\caption{\label{tab:restricted_bart} \black Results of restricted-area sampling on BART-Lab}
	\resizebox{\linewidth}{!}{%
		\begin{tabular}{c | c | c | c | c | c | c | c}
			\toprule
			\textbf{Sampling rate} & \textbf{Metrics} & Kriging & RME-GAN & {\black RME-GAN$^\dagger$} & LaPnP & {\black RadioDiff-Inverse} & RadioTrace \\
			\hline
			\multirow{4}{*}{1\%}
			& NMSE            & 0.0312 & 0.0391 & {\black \textbf{0.0092}} & 0.0352 & {\black 0.5326} & \underline{0.0175} \\
			& RMSE            & 0.0634 & 0.0798 & {\black \textbf{0.0469}} & 0.0624 & {\black 0.4496} & \underline{0.0504} \\
			& SSIM $\uparrow$ & 0.9089 & 0.9145 & {\black \underline{0.9210}} & 0.8969 & {\black 0.2133} & \textbf{0.9223} \\
			& PSNR $\uparrow$ & 24.60 & 21.99 & {\black \underline{25.59}} & 24.20 & {\black 6.94} & \textbf{26.16} \\
			\hline
			\multirow{4}{*}{5\%}
			& NMSE            & 0.0232 & 0.0281 & {\black \textbf{0.0081}} & 0.0217 & {\black 0.2879} & \underline{0.0136} \\
			& RMSE            & 0.0540 & 0.0703 & {\black \textbf{0.0393}} & 0.0508 & {\black 0.1930} & \underline{0.0418} \\
			& SSIM $\uparrow$ & 0.9213 & 0.9189 & {\black \underline{0.9264}} & 0.9219 & {\black 0.6225} & \textbf{0.9308} \\
			& PSNR $\uparrow$ & 25.31 & 23.09 & {\black \underline{26.11}} & 26.03 & {\black 14.29} & \textbf{27.31} \\
			\hline
			\multirow{4}{*}{10\%}
			& NMSE            & 0.0187 & 0.0204 & {\black \textbf{0.0077}} & 0.0155 & {\black 0.2389} & \underline{0.0126} \\
			& RMSE            & 0.0509 & 0.0687 & {\black \textbf{0.0376}} & 0.0480 & {\black 0.1603} & \underline{0.0398} \\
			& SSIM $\uparrow$ & 0.9277 & 0.9221 & {\black 0.9310} & \underline{0.9325} & {\black 0.7428} & \textbf{0.9345} \\
			& PSNR $\uparrow$ & 25.96 & 23.30 & {\black 26.51} & \underline{26.56} & {\black 16.04} & \textbf{27.63} \\
			\hline
			\multirow{4}{*}{20\%}
			& NMSE            & 0.0150 & 0.0182 & {\black \textbf{0.0076}} & 0.0141 & {\black 0.2090} & \underline{0.0114} \\
			& RMSE            & 0.0473 & 0.0684 & {\black \textbf{0.0373}} & 0.0459 & {\black 0.1479} & \underline{0.0384} \\
			& SSIM $\uparrow$ & 0.9348 & 0.9257 & {\black 0.9311} & \textbf{0.9379} & {\black 0.7971} & \underline{0.9368} \\
			& PSNR $\uparrow$ & 26.58 & 23.33 & {\black 26.77} & \underline{26.96} & {\black 17.44} & \textbf{27.97} \\
			\bottomrule
		\end{tabular}%
	}
\end{table}

{\black The quantitative
results are reported in Table~\ref{tab:restricted}. Compared with
Kriging, LaPnP, RME-GAN, and RadioDiff-Inverse, RadioTrace consistently
achieves lower RMSE and higher SSIM/PSNR across all sampling ratios. At the
1\% sampling ratio, RadioTrace reduces RMSE from 0.0958 to 0.0296 compared
with the default RME-GAN and improves PSNR from 20.37 dB to 31.95 dB. This
large gap suggests that a supervised model trained only under random sampling
can suffer from a clear sampling-pattern mismatch when directly applied to
restricted-area measurements.
We also include RME-GAN$^\dagger$. As expected, this stronger supervised baseline
substantially improves over the default RME-GAN and achieves the lowest NMSE
in Table~\ref{tab:restricted}. Nevertheless, RadioTrace still obtains
lower RMSE and higher SSIM/PSNR at all sampling ratios. 

The visual results in the last two rows of
Fig.~\ref{fig:results} further show that RME-GAN$^\dagger$ can recover
the intensity values within the restricted regions, but its reconstructed maps
may still blur or distort the underlying propagation structures. In contrast,
RadioTrace produces more geometry-consistent reconstructions around buildings
and unobserved regions.}

\begin{figure*}[htb]
	
	\centering
	\centerline{\includegraphics[width=18cm]{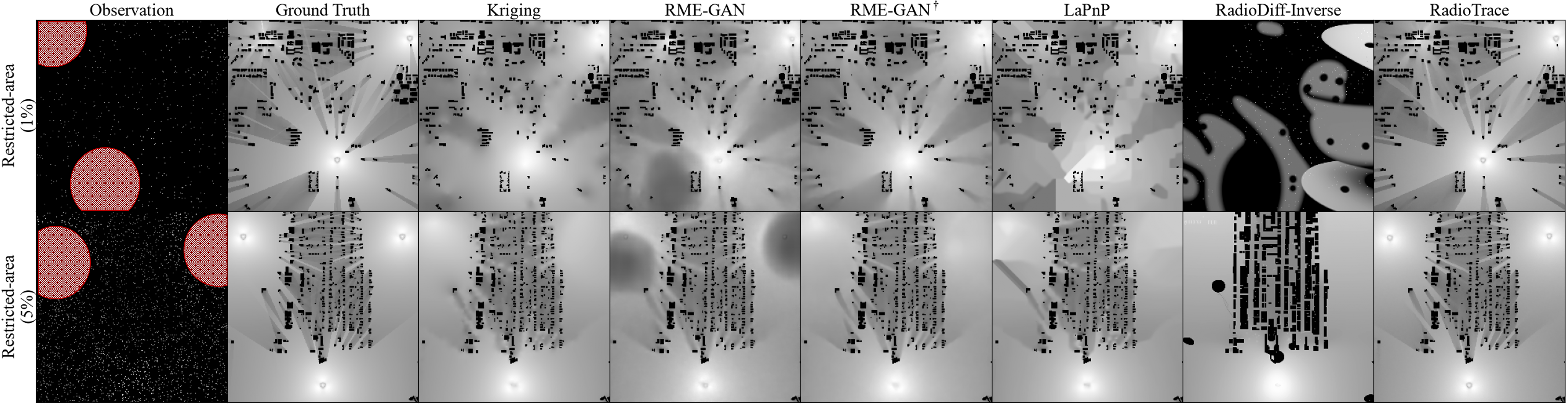}}
	
	\caption{\black Comparison of RM Estimation on different methods under \textbf{restricted-area sampling} with low sampling ratios (1\% and 5\%). The red regions correspond to restricted areas where no measurements are available.}
	\label{fig:results_bart}
\end{figure*}

\subsubsection{Results on BART-Lab dataset}

{\black
	To further assess the methods under more challenging conditions, we evaluate
	them on the BART-Lab dataset using the same restricted-area sampling setting. Compared
	with RadioMapSeer, BART-Lab contains finer building layouts and more complex
	multi-Tx propagation patterns. As reported in Table~\ref{tab:restricted_bart},
	RadioTrace achieves the highest PSNR across all sampling ratios and lower
	RMSE than the non-retrained baselines. The visual results in Fig.~\ref{fig:results_bart} show a similar trend.
	Kriging and LaPnP recover the coarse field but tend to oversmooth the
	unobserved regions, while RME-GAN introduces visible artifacts under the
	sampling-pattern shift. RadioDiff-Inverse shows evident degradation at low
	sampling ratios. RME-GAN$^\dagger$ remains competitive, but its results are
	still relatively blurry within restricted regions. In comparison, RadioTrace
	produces clearer large-scale propagation structures.}

\begin{table}[!t]
	\belowrulesep=0pt
	\aboverulesep=0pt
	\centering
	\caption{Ablation under \textbf{restricted-area} sampling at \textbf{1\%}. }
	\label{tab:ablation}
	\setlength{\tabcolsep}{4.5pt} 
	\renewcommand{\arraystretch}{1.1}
	\resizebox{\columnwidth}{!}{
		\begin{tabular}{ccc|cccc}
			\toprule
			\multicolumn{3}{c|}{\textbf{Components}} & \multicolumn{4}{c}{\textbf{Metrics}} \\
			\hline
			\makecell{PG\\K-means} & \makecell{Momentum\\GD} & \makecell{Best\\anchor} &
			NMSE & RMSE & SSIM $\uparrow$ & PSNR $\uparrow$ \\
			\hline
			\checkmark &              &              & 0.0316 & 0.0394 & 0.9367 & 30.00 \\
			\checkmark & \checkmark   &              & 0.0205 & 0.0330 & 0.9472 & 31.07 \\
			& \checkmark   & \checkmark   & 0.0264 & 0.0402 & 0.9419 & 29.97 \\
			\checkmark & \checkmark   & \checkmark   & \textbf{0.0174} & \textbf{0.0296} & \textbf{0.9518} & \textbf{31.95} \\
			\bottomrule
	\end{tabular}}
\end{table}

\begin{figure}[tb]
	
	\centering
	\centerline{\includegraphics[width=9cm]{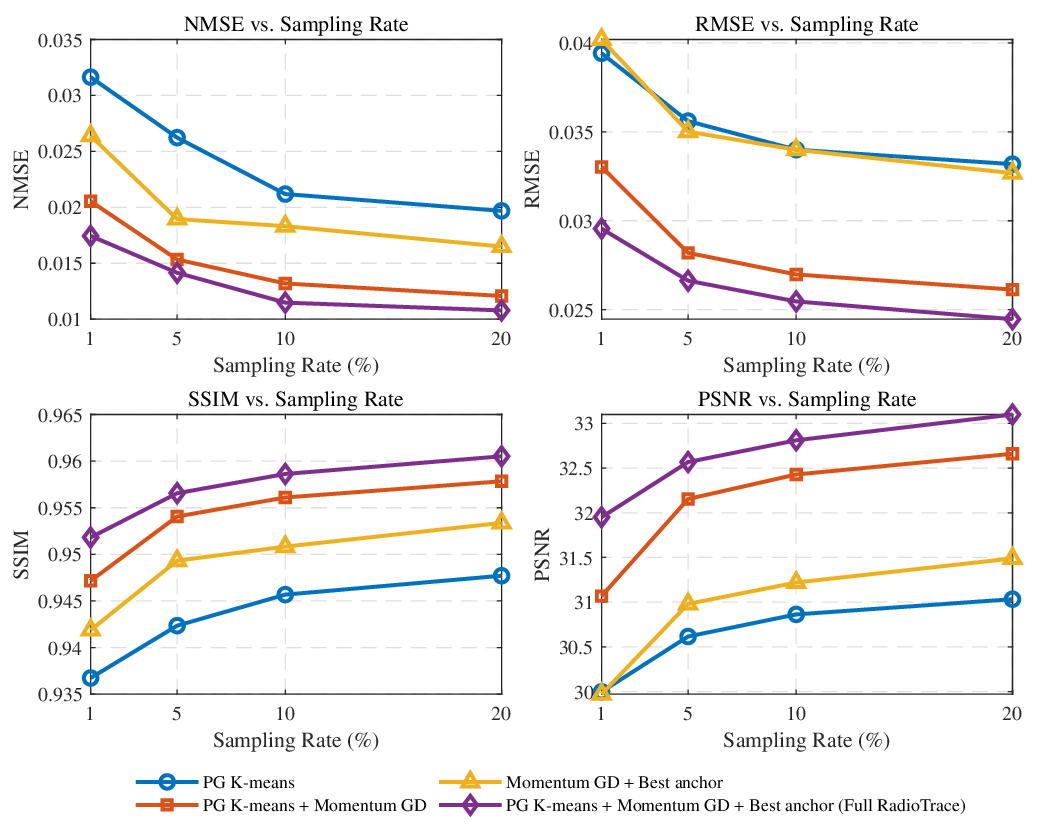}}
	
	\caption{Ablation under restricted-area sampling: metric–rate curves (NMSE, RMSE, SSIM, PSNR) versus sampling rate (1\%, 5\%, 10\%, 20\%) for four component configurations— PG K-means; PG K-means + Momentum GD; Momentum GD + Best anchor; and PG K-means + Momentum GD + Best anchor (Full RadioTrace).}
	\label{fig:ablation-line-2x2}
\end{figure}

\subsection{Ablation Studies}
Next, we conduct ablation studies for RadioTrace. Unless otherwise specified, all subsequent experiments are conducted under the restricted-area sampling setting. As summarized in Table~\ref{tab:ablation}, PG K-means alone provides a geometry-consistent starting point but plateaus under extreme sparsity; adding Momentum GD tightens errors (about 16\% RMSE reduction), indicating more effective assimilation of biased, sparse observations. Using Momentum GD with the best anchor without PG K-means remains limited in this regime, showing that late-stage stabilization cannot compensate for poor basin selection. The full configuration achieves the strongest performance, with RMSE about 25\% lower than PG K-means and 10\% below PG K-means + Momentum GD, demonstrating that the anchor adds value once the iterate is already close to a plausible geometry. {\black The metric curves in Fig.~\ref{fig:ablation-line-2x2} further show that the gain of the full model is consistent from 1\% to 20\% sampling.}


\subsection{Hyperparameter sensitivity}
All experiments in this subsection are conducted with a 1\% sampling rate. We report PSNR trends using the two plots in Fig.~\ref{fig:psnr-hparam}.

\subsubsection{Variance $\sigma^2$ of the Gaussian Tx map}
\begin{figure}[!t]
	\centering
	\subfloat[PSNR vs.\ $\sigma^2$ (log-x)]{
		\includegraphics[width=0.47\linewidth]{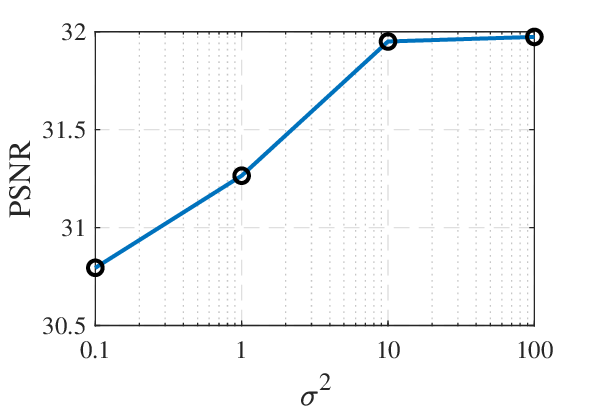}
	}\hfill
	\subfloat[PSNR vs.\ $\beta$]{
		\includegraphics[width=0.47\linewidth]{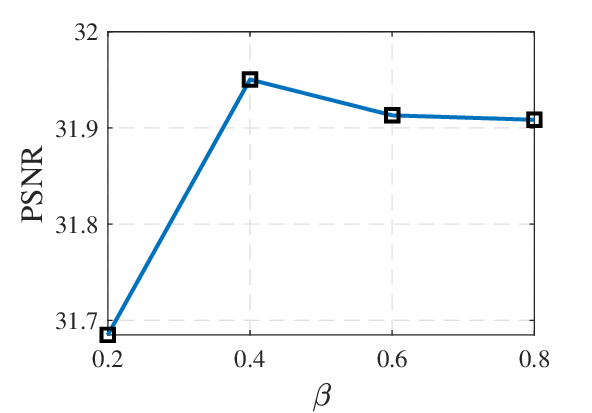}
	}
	\caption{PSNR sensitivity at 1\% sampling under restricted-area sampling.}
	\label{fig:psnr-hparam}
\end{figure}

Fig.~\ref{fig:psnr-hparam}(a) shows PSNR versus $\sigma^2$ on a logarithmic x-axis. PSNR rises from $\sigma^2{=}0.1$ to $10$ and then saturates near $\sigma^2{=}100$. A small $\sigma^2$ produces an overly sharp Tx prior that behaves like a near-delta seed. Under sparse and spatially biased observations, such a sharp seed is sensitive to small localization errors and to missing measurements near the true source, which leads to larger reconstruction error. Increasing $\sigma^2$ widens the prior and spreads energy over a neighborhood, which reduces sensitivity to missing samples and helps the optimizer enter a geometry-consistent basin. The gains diminish once the initialization is already inside a good basin, so further widening yields only small benefits. This explains the observed saturation and suggests a practical range $\sigma^2\in[10,100]$ for this regime.

\subsubsection{Momentum coefficient $\beta$ in GD}
Fig.~\ref{fig:psnr-hparam}(b) plots PSNR versus $\beta$. The curve has a shallow maximum around $\beta= 0.4$. With a very small $\beta$, updates rely mostly on the current noisy gradient, which slows progress and increases iterate jitter when measurements are scarce. With a large $\beta$, the velocity accumulates stale directions from earlier steps, which can overshoot near the solution and reduce late-stage accuracy. A mid-range momentum balances variance reduction and stability, allowing faster travel along consistent directions while still adapting at the end of optimization. The surface is relatively flat around the optimum, which indicates that the method is tolerant to moderate variation in $\beta$.

\begin{table}[!t]
	\centering
	\belowrulesep=0pt
	\aboverulesep=0pt
	\caption{Results under noisy observations in the restricted-area setting.}
	\label{tab:noisy-robust}
	\setlength{\tabcolsep}{6pt}
	\renewcommand{\arraystretch}{1.1}
	\begin{tabular}{c|c|cccc}
		\toprule
		\textbf{Std.\ (\(\sigma_n\))} & \textbf{Rate} & NMSE & RMSE & SSIM $\uparrow$ & PSNR $\uparrow$ \\
		\hline
		0 (no noise) & 1\% & 0.0174 & 0.0296 & 0.9518 & 31.95 \\
		0.01         & 1\% & 0.0178 & 0.0298 & 0.9512 & 31.86 \\
		0.10         & 1\% & 0.0221 & 0.0344 & 0.9455 & 30.79 \\
		\hline
		0 (no noise) & 5\% & 0.0141 & 0.0266 & 0.9566 & 32.57 \\
		0.01         & 5\% & 0.0138 & 0.0265 & 0.9567 & 32.61 \\
		0.10         & 5\% & 0.0131 & 0.0280 & 0.9547 & 32.14 \\
		\bottomrule
	\end{tabular}
\end{table}

\subsection{Results under noisy observations}
We assess robustness to measurement noise, which typically stems from device errors, by adding zero-mean Gaussian perturbations with standard deviation $\sigma_n$ after normalizing the dB maps to $[-1, 1]$. Experiments are conducted at 1\% and 5\% sampling rates. As summarized in Table~\ref{tab:noisy-robust}, mild noise ($\sigma_n=0.01$) stays close to the clean baseline at both rates, while stronger noise ($\sigma_n=0.10$) causes a moderate, smooth drop in PSNR and a small rise in RMSE. 

\begin{figure}[tb]
	
	\centering
	\centerline{\includegraphics[width=9cm]{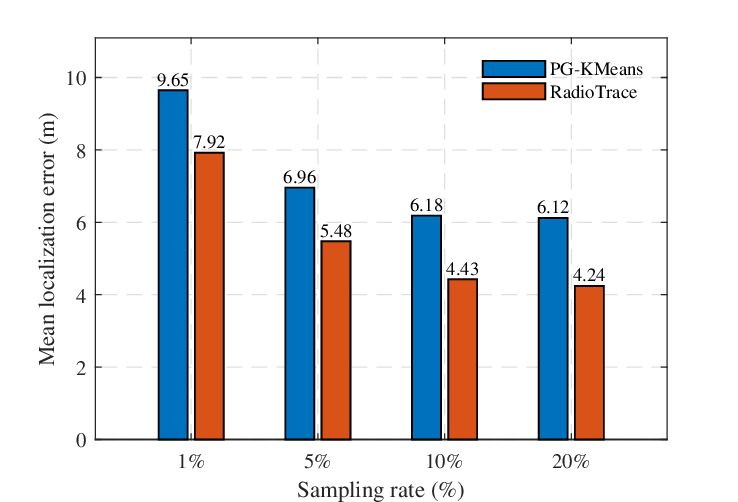}}
	
	\caption{Transmitter localization accuracy under restricted-area sampling.}
	\label{fig:txloc-bar}
\end{figure}

\subsection{Localization accuracy}
In this experiment, we also evaluate Tx localization accuracy results under the restricted-area sampling mode, though our main goal is to recover RM. In our pipeline, PG-KMeans is the initialization in the proposed RadioTrace and provides the initialized Tx positions, while RadioTrace denotes the whole framework containing PG-KMeans and Tx refinement. As shown in Fig.~\ref{fig:txloc-bar}, the update consistently improves over the initialization at 1\%, 5\%, 10\%, and 20\% sampling. The gap is visible at low sampling rates and grows slightly with more observations, whereas the initialization curve tends to plateau at higher rates. These results indicate that the update stage, informed by the concurrent map reconstruction, adds clear value beyond the PG-KMeans initializer and yields lower mean localization error across sampling levels.

\begin{table}[!t]
	\centering
	\belowrulesep=0pt
	\aboverulesep=0pt
	\caption{Extreme low sampling rate results under restricted-area sampling on a 256$\times$256 grid (65{,}536 pixels).}
	\label{tab:extreme-low-rate}
	\setlength{\tabcolsep}{5pt}
	\renewcommand{\arraystretch}{1.1}
	\begin{tabular}{c|c|cccc}
		\toprule
		\textbf{Rate} & \textbf{Points} & NMSE & RMSE & SSIM $\uparrow$ & PSNR $\uparrow$\\
		\hline
		0.10\% & $\approx$66 & 0.0435 & 0.0435 & 0.9307 & 29.84 \\
		0.05\% & $\approx$33 & 0.0675 & 0.0592 & 0.9087 & 27.40 \\
		\bottomrule
	\end{tabular}
\end{table}

\subsection{Extreme low sampling rate performance}
We evaluate RadioTrace at extremely low sampling rates 0.1\% and 0.05\%. On a 256×256 grid (65,536 pixels), this corresponds to roughly 66 and 33 measurements, respectively. As shown in Table~\ref{tab:extreme-low-rate}, at sampling rate of 0.1\%, we obtain NMSE 0.0435, RMSE 0.0435, SSIM 0.9307, and PSNR 29.84 dB; the mean Tx error improves from 24.78 m at initialization to 20.45 m after refinement ($\approx$ 4.33 m reduction). At 0.05\%, performance drops as expected (NMSE 0.0675, RMSE 0.0592, SSIM 0.9087, PSNR 27.40 dB), while localization still improves from 41.28 m to 33.88 m ($\approx$ 7.40 m reduction). {\black These results show that RadioTrace remains effective even when only very limited measurements are available.}

\subsection{Runtime and complexity}
We study the cost–quality tradeoff on one A800 GPU by varying the number of sampling steps $T\in\{100,50,10\}$ at 1\% observations. The reported rate is an approximate, measured end-to-end throughput that includes typical overheads such as data movement and light pre/post-processing. As $T$ decreases, the rate increases while accuracy drops smoothly; a practical choice is $T=50$, which preserves most of the quality of $T=100$ at roughly double the measured rate, while $T=10$ offers a large speedup with moderate loss. In practice, RadioTrace can also be paired with fast samplers such as denoising diffusion implicit models (DDIM) to reduce the required steps, and the backbone capacity can be chosen to match latency and memory constraints so that lighter models with accelerated sampling favor throughput, whereas heavier models with moderate steps favor accuracy.

\begin{table}[!t]
	\centering
	\belowrulesep=0pt
	\aboverulesep=0pt
	\caption{Cost–quality tradeoff at 1\% sampling under restricted-area sampling}
	\label{tab:runtime}
	\setlength{\tabcolsep}{6pt}
	\renewcommand{\arraystretch}{1.1}
	\begin{tabular}{c|cccc|c}
		\toprule
		\textbf{$T$} & NMSE & RMSE & SSIM $\uparrow$ & PSNR $\uparrow$ & rate (maps/s) \\
		\hline
		100 & 0.0174 & 0.0296 & 0.9518 & 31.95 & 0.0618 \\
		50  & 0.0186 & 0.0306 & 0.9497 & 31.70 & 0.1187 \\
		10  & 0.0222 & 0.0327 & 0.9468 & 31.28 & 0.5832 \\
		\bottomrule
	\end{tabular}
\end{table}

We also report the average inference time per RM reconstruction on the same device. Interpolation methods like Kriging are the fastest ($<0.1$ s), making them attractive for strictly resource-constrained scenarios; however, their accuracy collapses under non-uniform or restricted-area sampling. RME-GAN requires only 0.2 s via a single forward pass, but it operates under a much more favorable formulation that strictly requires known Tx locations. Furthermore, its brief runtime excludes substantial offline training (12 hours) and the necessity for retraining under sampling shifts. In contrast, RadioTrace and LaPnP solve an iterative inverse problem at deployment without any Tx knowledge. With $T{=}10$ reverse steps, RadioTrace takes 1.7 s, which is significantly more efficient than LaPnP (25 s). {\black RadioDiff-Inverse is also computationally demanding, requiring about 198 s per reconstruction due to its diffusion-based posterior sampling procedure.} Overall, RadioTrace offers a highly favorable accuracy-deployability tradeoff for performance-oriented settings.

{\black 
\subsection{Scalability}
We evaluate scalability with respect to the number of Tx by varying $R \in \{1,2,3,4,5\}$ with  1\% sampling rate, while keeping the environment, map resolution, and measurement protocol fixed. 
As shown in Fig.~\ref{fig:scalability_tx}, as the number of Tx increases, the RMSE gradually rises while the PSNR consistently decreases, indicating that the reconstruction task becomes more challenging in more complex multi-Tx scenarios. Nevertheless, the overall degradation remains moderate, which demonstrates that the proposed method still maintains reasonable scalability.

As the number of Tx $R$ increases, the only additional computations introduced are the generation of the $R$ soft Gaussian Tx maps and the calculation of their gradients with respect to the Tx coordinates. Since these are simple, low-dimensional mathematical operations, their overhead is strictly negligible compared to the heavy neural network evaluations of the diffusion backbone. Therefore, an increase in $R$ adds practically zero computational burden to the overall inference pipeline.

\begin{figure}[tb]
	
	\centering
	\centerline{\includegraphics[width=7cm]{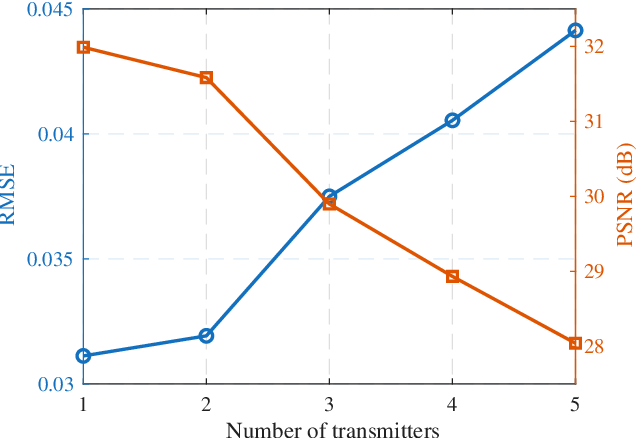}}
	
	\caption{\black Scalability of number of transmitters.}
	\label{fig:scalability_tx}
\end{figure}
}

\begin{figure}[t]
	\centering
	\includegraphics[width=0.95\linewidth]{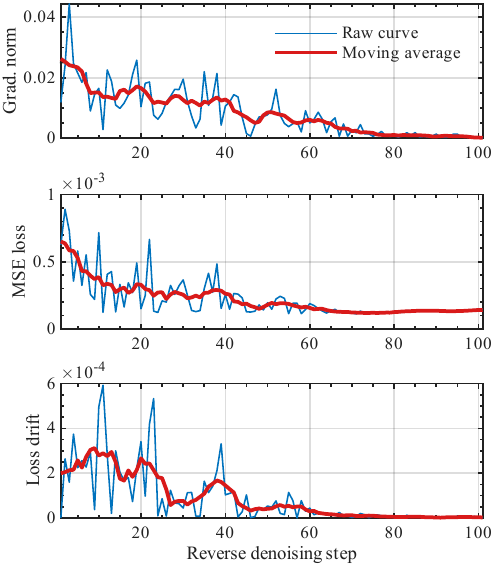}
	\caption{\black Empirical behavior of Tx-coordinate refinement during reverse denoising.}
	\label{fig:denoising_curves}
\end{figure}

{\black
\subsection{Empirical Behavior of Tx Refinement During Denoising} \label{sec:refine}

We further record the Tx-coordinate refinement trajectory of a representative test sample during reverse denoising. 
As shown in Fig.~\ref{fig:denoising_curves}, the three panels from top to bottom report the gradient norm of the refinement loss, the MSE loss, and the absolute inter-step loss variation, respectively. 
The blue curves denote the raw recorded values, while the red curves denote the moving-average trends.
It can be observed that the curves fluctuate more noticeably in the early denoising stage, where the generated RM and Tx locations are still uncertain. 
As denoising proceeds, the moving-average curves gradually decrease and become smoother, especially for the gradient norm and the absolute inter-step loss variation. 
This indicates that the Tx-coordinate refinement process becomes more stable in later denoising steps, which is consistent with the controlled objective drift and gradient perturbation assumed in the stability analysis.
}

\section{Conclusion}

In this paper, we presented RadioTrace, a transmitter-aware diffusion framework for sparse-measurement RM estimation with unknown Tx locations. By treating Tx coordinates as explicit physical variables and refining them within the reverse diffusion process, RadioTrace enables sparse RSS measurements to update both the reconstructed RM and the Tx-conditioned generative process. A propagation-guided K-means initializer is introduced to provide geometry-consistent Tx seeds, and the stability of the Tx-coordinate refinement process is analyzed under diffusion-induced randomness. Experiments under different sampling settings demonstrate the effectiveness of RadioTrace for joint RM reconstruction and Tx localization. These results suggest that incorporating source-related physical variables into diffusion-based inverse reconstruction can improve the practical applicability of generative priors for radio-environment mapping.

{\black 

\appendix

\subsection{Proof of Theorem~\ref{thm:main}}\label{app:thm}

Let $g_k:=\nabla \bar L_k(\Omega_k)$ and $G_k=g_k+\zeta_k$. Define
$m_k:=\mathbb{E}[\zeta_k\mid\mathcal{F}_k]$ and
$\tilde\zeta_k:=\zeta_k-m_k$, so that
$\mathbb{E}[\tilde\zeta_k\mid\mathcal{F}_k]=0$,
$\mathbb{E}[\|\tilde\zeta_k\|^2\mid\mathcal{F}_k]\le \sigma_k^2$, and
$\zeta_k=m_k+\tilde\zeta_k$. The update rule is reexpressed as
\begin{equation}
	\begin{aligned}
		v_{k+1}&=\beta v_k+(1-\beta)(g_k+m_k+\tilde\zeta_k),\\
		\Omega_{k+1}&=\Omega_k-\eta_k v_{k+1}.
	\end{aligned}
	\label{eq:app:update}
\end{equation}

By the $M_k$-smoothness of $\bar L_k$ and $\Omega_{k+1}-\Omega_k=-\eta_k v_{k+1}$,
\begin{equation}
	\bar L_k(\Omega_{k+1})
	\le
	\bar L_k(\Omega_k)
	-\eta_k\langle g_k,v_{k+1}\rangle
	+\frac{M_k\eta_k^2}{2}\|v_{k+1}\|^2.
	\label{eq:app:smooth}
\end{equation}

Let $e_k:=m_k+\tilde\zeta_k$. From \eqref{eq:app:update},
$v_{k+1}-\beta v_k=(1-\beta)(g_k+e_k)$. Squaring this identity and eliminating
$2\beta\langle v_k,v_{k+1}\rangle$ gives
\begin{equation}
	2(1-\beta)\langle g_k,v_{k+1}\rangle
	=
	\|v_{k+1}\|^2-\beta^2\|v_k\|^2+(1-\beta)^2\|g_k\|^2+R_k,
	\label{eq:app:momentum}
\end{equation}
where
\begin{equation}
	R_k:=(1-\beta)^2\!\left(\|e_k\|^2+2\langle g_k,e_k\rangle\right)
	-2(1-\beta)\langle e_k,v_{k+1}\rangle .
	\label{eq:app:Rdef}
\end{equation}
Substituting \eqref{eq:app:momentum} into \eqref{eq:app:smooth} yields
\begin{equation}
	\begin{aligned}
		\bar L_k(\Omega_{k+1})
		\le {}&
		\bar L_k(\Omega_k)
		-\frac{\eta_k(1-\beta)}{2}\|g_k\|^2
		+\frac{\eta_k\beta^2}{2(1-\beta)}\|v_k\|^2 \\
		&-\Bigl(\frac{\eta_k}{2(1-\beta)}-\frac{M_k\eta_k^2}{2}\Bigr)\|v_{k+1}\|^2
		-\frac{\eta_k}{2(1-\beta)}R_k .
	\end{aligned}
	\label{eq:app:descent}
\end{equation}

By \eqref{eq:app:Rdef} and Young's inequality, for any $\rho,\varepsilon>0$,
\begin{equation}
	|R_k|
	\le
	\rho(1-\beta)^2\|g_k\|^2
	+\varepsilon\|v_{k+1}\|^2
	+C\|e_k\|^2,
	\label{eq:app:Rbound}
\end{equation}
where $C:=(1-\beta)^2(1+\rho^{-1}+\varepsilon^{-1})$. Moreover,
\[
\mathbb{E}[\|e_k\|^2\mid\mathcal{F}_k]
=
\|m_k\|^2+\mathbb{E}[\|\tilde\zeta_k\|^2\mid\mathcal{F}_k]
\le
\delta_k^2+\sigma_k^2
\le
C_0(\delta_k+\sigma_k^2),
\]
for some constant $C_0$ independent of $k$, where the last step uses $\delta_k\le \bar\delta$. Hence,
\[
\mathbb{E}[|R_k|\mid\mathcal{F}_k]
\le
c_1\|g_k\|^2
+\varepsilon\,\mathbb{E}[\|v_{k+1}\|^2\mid\mathcal{F}_k]
+C_1(\delta_k+\sigma_k^2),
\]
with $c_1:=\rho(1-\beta)^2$ and some constant $C_1$. Taking conditional expectation in
\eqref{eq:app:descent} and using the above bound, we obtain
\begin{equation}
	\begin{aligned}
		\mathbb{E}[\bar L_k(\Omega_{k+1})\mid\mathcal{F}_k]
		\le {}&
		\bar L_k(\Omega_k)
		-\eta_k\Bigl(\frac{1-\beta}{2}-\frac{c_1}{2(1-\beta)}\Bigr)\|g_k\|^2 \\
		&+\frac{\eta_k\beta^2}{2(1-\beta)}\|v_k\|^2
		-A_k
		\mathbb{E}[\|v_{k+1}\|^2\mid\mathcal{F}_k] \\
		&+C_2\eta_k(\delta_k+\sigma_k^2)
	\end{aligned}
	\label{eq:app:cond-descent}
\end{equation}
for some constant $C_2$ and $
A_k:=\frac{\eta_k}{2(1-\beta)}-\frac{M_k\eta_k^2}{2}
-\frac{\eta_k\varepsilon}{2(1-\beta)}.
$
Choosing $\rho\in(0,1)$ and defining
$c_2:=\frac{1-\beta}{2}(1-\rho)>0$, \eqref{eq:app:cond-descent} becomes
\begin{equation}
	\begin{aligned}
		\mathbb{E}[\bar L_k(\Omega_{k+1})\mid\mathcal{F}_k]
		\le {}&
		\bar L_k(\Omega_k)-c_2\eta_k\|g_k\|^2
		+\frac{\eta_k\beta^2}{2(1-\beta)}\|v_k\|^2 \\
		&-A_k
		\mathbb{E}[\|v_{k+1}\|^2\mid\mathcal{F}_k]+C_2'\eta_k(\delta_k+\sigma_k^2).
	\end{aligned}
	\label{eq:app:cond-descent2}
\end{equation}
for some constant $C_2'$. Define the Lyapunov function
\[
\Phi_k:=\bar L_k(\Omega_k)+\lambda_k\|v_k\|^2,
\qquad
\lambda_k:=\frac{\eta_k\beta^2}{2(1-\beta)}.
\]
Using the drift condition
$\bar L_{k+1}(\Omega_{k+1})\le \bar L_k(\Omega_{k+1})+\bar d_k$ together with
\eqref{eq:app:cond-descent2}, we obtain
\begin{equation}
	\begin{aligned}
		\mathbb{E}[\Phi_{k+1}\mid\mathcal{F}_k]
		\le {}&
		\Phi_k-c_2\eta_k\|g_k\|^2+\bar d_k+C_2'\eta_k(\delta_k+\sigma_k^2) \\
		&-\Bigl(A_k-\lambda_{k+1}\Bigr)
		\mathbb{E}[\|v_{k+1}\|^2\mid\mathcal{F}_k].
	\end{aligned}
	\label{eq:app:lyapunov}
\end{equation}
Since $\eta_k$ is nonincreasing and
$\lambda_{k+1}\le \eta_k\beta^2/[2(1-\beta)]$, 
\begin{equation}
	\begin{aligned}
		A_k-\lambda_{k+1} &= \frac{\eta_k}{2(1-\beta)}-\frac{M_k\eta_k^2}{2}
		-\frac{\eta_k\varepsilon}{2(1-\beta)} - \lambda_{k+1} \\
		&\ge
		\eta_k\Bigl(
		\frac{1+\beta}{2}-\frac{\varepsilon}{2(1-\beta)}-\frac{M_k\eta_k}{2}
		\Bigr).
	\end{aligned}
	\label{eq:app:coef}
\end{equation}
Choose $\varepsilon>0$ sufficiently small such that
$\frac{1+\beta}{2}-\frac{\varepsilon}{2(1-\beta)}>0$.
Since $\eta_kM_k\to0$, there exist $k_0$ and $c_3>0$ such that the coefficient
in \eqref{eq:app:coef} is at least $c_3\eta_k$ for all $k\ge k_0$. Hence the
last term in \eqref{eq:app:lyapunov} is nonpositive for all sufficiently large
$k$, and thus for $k\ge k_0$,
\begin{equation}
	\mathbb{E}[\Phi_{k+1}\mid\mathcal{F}_k]
	\le
	\Phi_k-c_2\eta_k\|g_k\|^2+\bar d_k+C_2'\eta_k(\delta_k+\sigma_k^2).
	\label{eq:app:recursion}
\end{equation}

Taking full expectation and summing \eqref{eq:app:recursion} from $k_0$ to $K$
gives
\begin{equation}
	\begin{aligned}
		\mathbb{E}[\Phi_{K+1}]
		\le {}&
		\mathbb{E}[\Phi_{k_0}]
		-c_2\sum_{k=k_0}^{K}\eta_k\,\mathbb{E}[\|g_k\|^2]
		+\sum_{k=k_0}^{K}\bar d_k \\
		&+C_2'\sum_{k=k_0}^{K}\eta_k(\delta_k+\sigma_k^2).
	\end{aligned}
	\label{eq:app:sum}
\end{equation}
Since $\bar L_k(\Omega_k)\ge0$ and $\lambda_k\|v_k\|^2\ge0$, we have
$\Phi_k\ge0$. Using Assumption~2 and letting $K\to\infty$ in \eqref{eq:app:sum},
we conclude that
\begin{equation}
	\sum_{k=0}^{\infty}\eta_k\,\mathbb{E}[\|g_k\|^2]<\infty.
	\label{eq:app:summability}
\end{equation}
Equivalently,
\begin{equation}
	\sum_{k=0}^{\infty}\eta_k\,
	\mathbb{E}\!\left[\|\nabla \bar L_k(\Omega_k)\|^2\right]<\infty.
	\label{eq:app:summability2}
\end{equation}
Notice that  $\sum_{k=0}^{\infty}\eta_k=\infty$ by Assumption~\ref{ass:steps}, it follows that
\[
\liminf_{k\to\infty}\mathbb{E}[\|\nabla \bar L_k(\Omega_k)\|^2]=0.
\]
This completes the proof. 
%

}

\bibliographystyle{IEEEtran}
\bibliography{refs}

@ARTICLE{Levie2021RadioUNet,
  author={Levie, Ron and Yapar, Cagkan and Kutyniok, Gitta and Caire, Giuseppe},
  journal={IEEE Transactions on Wireless Communications}, 
  title={{RadioUNet}: Fast Radio Map Estimation With Convolutional Neural Networks}, 
  year={2021},
  volume={20},
  number={6},
  pages={4001-4015},
  doi={10.1109/TWC.2021.3054977}}

@ARTICLE{Wang2024RadioDiff,
  author={Wang, Xiucheng and Tao, Keda and Cheng, Nan and Yin, Zhisheng and Li, Zan and Zhang, Yuan and Shen, Xuemin},
  journal={IEEE Transactions on Cognitive Communications and Networking}, 
  title={{RadioDiff}: An Effective Generative Diffusion Model for Sampling-Free Dynamic Radio Map Construction}, 
  year={2024},
  volume={},
  number={},
  pages={1-1},
  doi={10.1109/TCCN.2024.3504489}}

@ARTICLE{Romero2022Radio,
  author={Romero, Daniel and Kim, Seung-Jun},
  journal={IEEE Signal Processing Magazine}, 
  title={Radio Map Estimation: A data-driven approach to spectrum cartography}, 
  year={2022},
  volume={39},
  number={6},
  pages={53-72},
  doi={10.1109/MSP.2022.3200175}}

@ARTICLE{Lee2012Voronoi,
  author={Lee, Minkyu and Han, Dongsoo},
  journal={IEEE Communications Letters}, 
  title={Voronoi Tessellation Based Interpolation Method for {Wi-Fi} Radio Map Construction}, 
  year={2012},
  volume={16},
  number={3},
  pages={404-407},
  doi={10.1109/LCOMM.2012.020212.111992}}

@ARTICLE{Teganya2022DeepCA,
  author={Teganya, Yves and Romero, Daniel},
  journal={IEEE Transactions on Wireless Communications}, 
  title={Deep Completion Autoencoders for Radio Map Estimation}, 
  year={2022},
  volume={21},
  number={3},
  pages={1710-1724},
  doi={10.1109/TWC.2021.3106154}}

@ARTICLE{Zhang2023RMEGAN,
  author={Zhang, Songyang and Wijesinghe, Achintha and Ding, Zhi},
  journal={IEEE Internet of Things Journal}, 
  title={{RME-GAN}: A Learning Framework for Radio Map Estimation Based on Conditional Generative Adversarial Network}, 
  year={2023},
  volume={10},
  number={20},
  pages={18016-18027},
  doi={10.1109/JIOT.2023.3278235}}

@article{jia2025rmdm,
  title={{RMDM}: Radio Map Diffusion Model with Physics Informed},
  author={Jia, Haozhe and Chen, Wenshuo and Huang, Zhihui and Xiao, Hongru and Jia, Nanqian and Wu, Keming and Lai, Songning and Yue, Yutao},
  journal={arXiv preprint arXiv:2501.19160},
  year={2025}
}

@ARTICLE{Xu2025Radio,
  author={Xu, Le and Cheng, Lei and Chen, Junting and Pu, Wenqiang and Fu, Xiao},
  journal={IEEE Transactions on Signal Processing}, 
  title={Radio Map Estimation via Latent Domain Plug-and-Play Denoising}, 
  year={2026},
  volume={},
  number={},
  pages={1-14},
  doi={10.1109/TSP.2025.3650699}}

@article{bengio2013estimating,
  title={Estimating or propagating gradients through stochastic neurons for conditional computation},
  author={Bengio, Yoshua and L{\'e}onard, Nicholas and Courville, Aaron},
  journal={arXiv preprint arXiv:1308.3432},
  year={2013}
}

@ARTICLE{yun2015ray,
  author={Yun, Zhengqing and Iskander, Magdy F.},
  journal={IEEE Access}, 
  title={Ray Tracing for Radio Propagation Modeling: Principles and Applications}, 
  year={2015},
  volume={3},
  number={},
  pages={1089-1100}}

@ARTICLE{bi2019eng,
  author={Bi, Suzhi and Lyu, Jiangbin and Ding, Zhi and Zhang, Rui},
  journal={IEEE Wireless Communications}, 
  title={Engineering Radio Maps for Wireless Resource Management}, 
  year={2019},
  volume={26},
  number={2},
  pages={133-141}}

@ARTICLE{chen2024optimal,
  author={Chen, Yujing and Yang, Dingcheng and Xiao, Lin and Wu, Fahui and Xu, Yu},
  journal={IEEE Transactions on Vehicular Technology}, 
  title={Optimal Trajectory Design for Unmanned Aerial Vehicle Cargo Pickup and Delivery System Based on Radio Map}, 
  year={2024},
  volume={73},
  number={8},
  pages={11706-11718}}

@INPROCEEDINGS{zhang2024fast,
  author={Zhang, Zezhong and Zhu, Guangxu and Chen, Junting and Cui, Shuguang},
  booktitle={2024 IEEE International Conference on Communications Workshops (ICC Workshops)}, 
  title={Fast and Accurate Cooperative Radio Map Estimation Enabled by {GAN}}, 
  year={2024},
  volume={},
  number={},
  pages={1641-1646}}

@ARTICLE{luo2025denoising,
  author={Luo, Xuanhao and Li, Zhizhen and Peng, Zhiyuan and Chen, Mingzhe and Liu, Yuchen},
  journal={IEEE Transactions on Cognitive Communications and Networking}, 
  title={Denoising Diffusion Probabilistic Model for Radio Map Estimation in Generative Wireless Networks}, 
  year={2025},
  volume={11},
  number={2},
  pages={751-763},}

@ARTICLE{wang2025radiodiff,
  author={Wang, Xiucheng and Fang, Zhongsheng and Cheng, Nan and Sun, Ruijin and Zhou, Haibo and Su, Zhou and Li, Zan and Shen, Xuemin},
  journal={IEEE Transactions on Wireless Communications}, 
  title={{RadioDiff-Inverse}: Diffusion Enhanced Bayesian Inverse Estimation for ISAC Radio Map Construction}, 
  year={2026},
  volume={25},
  number={},
  pages={14611-14626}}

@INPROCEEDINGS{yang2025radiotrace,
  author={Yang, Liu and Li, Qiang and Cao, Zhuo and Lin, Jingran},
  booktitle={2025 IEEE 35th International Workshop on Machine Learning for Signal Processing (MLSP)}, 
  title={{RadioTrace}: Bridging Diffusion Priors and RSS Measurements for Accurate Radio Map Estimation}, 
  year={2025}}

@inproceedings{ronneberger2015u,
  title={U-net: Convolutional networks for biomedical image segmentation},
  author={Ronneberger, Olaf and Fischer, Philipp and Brox, Thomas},
  booktitle={International Conference on Medical image computing and computer-assisted intervention},
  pages={234--241},
  year={2015},
  organization={Springer}
}

@inproceedings{liu2021swin,
  title={Swin transformer: Hierarchical vision transformer using shifted windows},
  author={Liu, Ze and Lin, Yutong and Cao, Yue and Hu, Han and Wei, Yixuan and Zhang, Zheng and Lin, Stephen and Guo, Baining},
  booktitle={Proceedings of the IEEE/CVF international conference on computer vision},
  pages={10012--10022},
  year={2021}
}

@inproceedings{NIPS2017_7a98af17,
 author = {van den Oord, Aaron and Vinyals, Oriol and kavukcuoglu, koray},
 booktitle = {Advances in Neural Information Processing Systems},
 editor = {I. Guyon and U. Von Luxburg and S. Bengio and H. Wallach and R. Fergus and S. Vishwanathan and R. Garnett},
 pages = {},
 publisher = {Curran Associates, Inc.},
 title = {Neural Discrete Representation Learning},
 volume = {30},
 year = {2017}
}

@ARTICLE{zeng2024tutorial,
  author={Zeng, Yong and Chen, Junting and Xu, Jie and Wu, Di and Xu, Xiaoli and Jin, Shi and Gao, Xiqi and Gesbert, David and Cui, Shuguang and Zhang, Rui},
  journal={IEEE Communications Surveys \& Tutorials}, 
  title={A Tutorial on Environment-Aware Communications via Channel Knowledge Map for 6G}, 
  year={2024},
  volume={26},
  number={3},
  pages={1478-1519}}

@ARTICLE{zhao2025imnet,
  author={Zhao, Le and Fei, Zesong and Wang, Xinyi and Huang, Jingxuan and Li, Yuan and Zhang, Yan},
  journal={IEEE Wireless Communications Letters}, 
  title={{IMNet}: Interference-Aware Channel Knowledge Map Construction and Localization}, 
  year={2025},
  volume={14},
  number={3},
  pages={856-860}}

@article{fu2025ckmdiff,
  title={{CKMDiff}: A generative diffusion model for CKM construction via inverse problems with learned priors},
  author={Fu, Shen and Zeng, Yong and Wu, Zijian and Wu, Di and Jin, Shi and Wang, Cheng-Xiang and Gao, Xiqi},
  journal={arXiv preprint arXiv:2504.17323},
  year={2025}
}

@ARTICLE{zhang2024physics,
  author={Zhang, Songyang and Choi, Brian and Ouyang, Feng and Ding, Zhi},
  journal={IEEE Communications Magazine}, 
  title={Physics-Inspired Machine Learning for Radiomap Estimation: Integration of Radio Propagation Models and Artificial Intelligence}, 
  year={2024},
  volume={62},
  number={8},
  pages={155-161},
  doi={10.1109/MCOM.001.2300782}}

\begin{IEEEbiography}[{\includegraphics[width=1in,height=1.25in,clip,keepaspectratio]{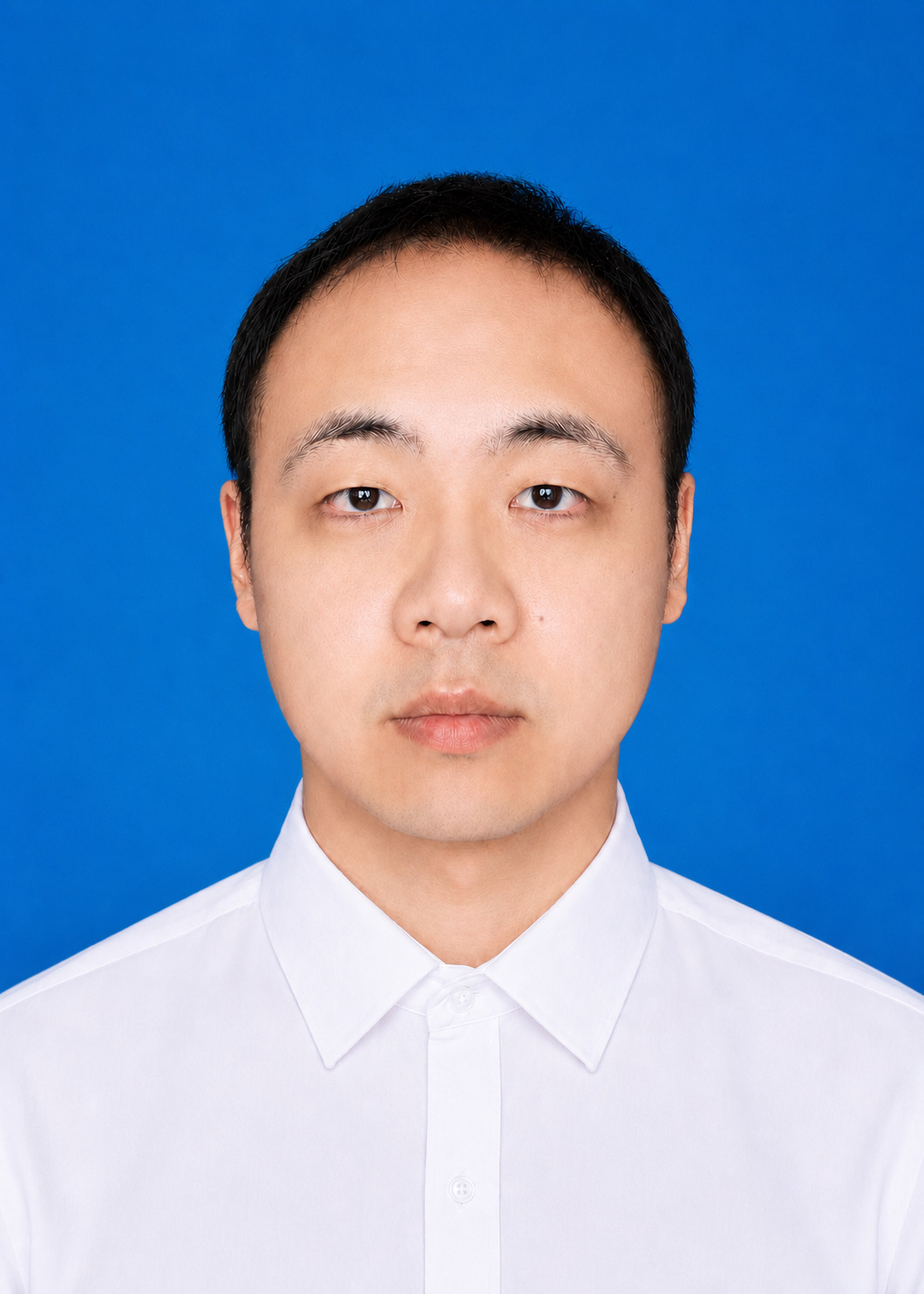}}]	{Liu Yang} received the B.Eng degree in 2020 from the University of Electronic Science and Technology of China (UESTC), Chengdu, China, where he is currently working toward the PhD degree with the School of Information and Communication Engineering. His current research interests include signal processing, radio frequency fingerprint identification and generative models.
\end{IEEEbiography}

\begin{IEEEbiography}[{\includegraphics[width=1in,height=1.25in,clip,keepaspectratio]{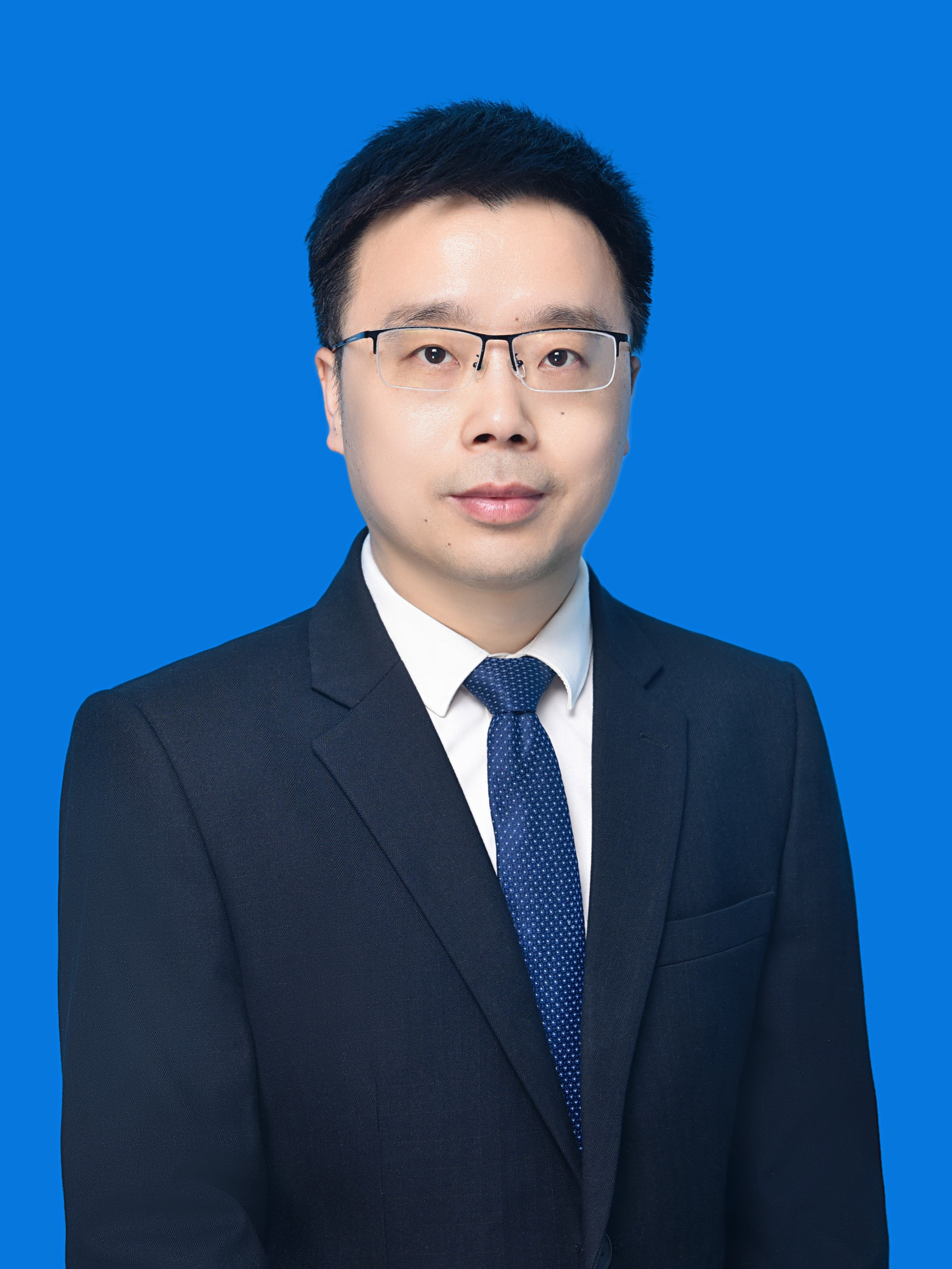}}]	{Qiang Li} received the B.Eng. and M.Phil. degrees in Communication and Information Engineering from University of Electronic Science and Technology of China (UESTC), Chengdu, China, and the Ph.D. degree in Electronic Engineering from the Chinese University of Hong Kong (CUHK), Hong Kong, in 2005, 2008, and 2012, respectively. He was a  Visiting Scholar with the University of Minnesota and Research Associate with the Department of Electronic Engineering and the Department of Systems Engineering and Engineering Management, CUHK. Since November 2013, he has been with the School of Information and Communication Engineering, UESTC, where he is currently a Professor. His recent research interests focus on machine learning and intelligent signal processing	in wireless communications.	He received a Best Paper Award of IEEE PIMRC 2016, and the Best Paper Award of the IEEE Signal Processing Letters 2016.
\end{IEEEbiography}

\vspace{-20pt}

\begin{IEEEbiography}[{\includegraphics[width=1in,height=1.25in,clip,keepaspectratio]{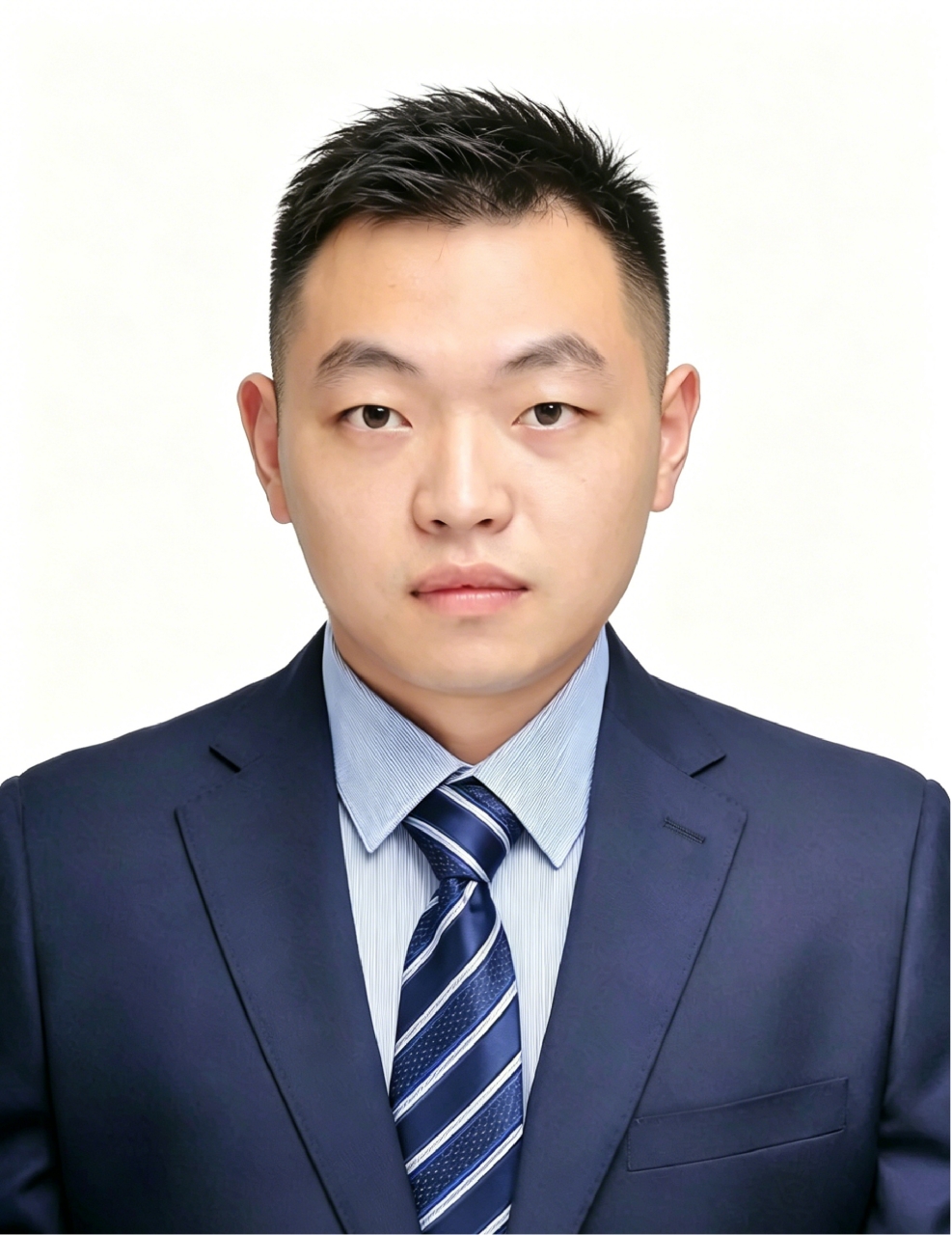}}]	{Zhuo Cao} received the B.S. degree from Anhui Normal University, Wuhu, Anhui, China, and is currently pursuing the Master's degree with the School of Information and Communication Engineering, University of Electronic Science and Technology of China (UESTC), Chengdu, Sichuan, China. His research interests include wireless communications, signal processing, deep learning, and radio map construction.
\end{IEEEbiography}

\begin{IEEEbiography}[{\includegraphics[width=1in,height=1.25in,clip,keepaspectratio]{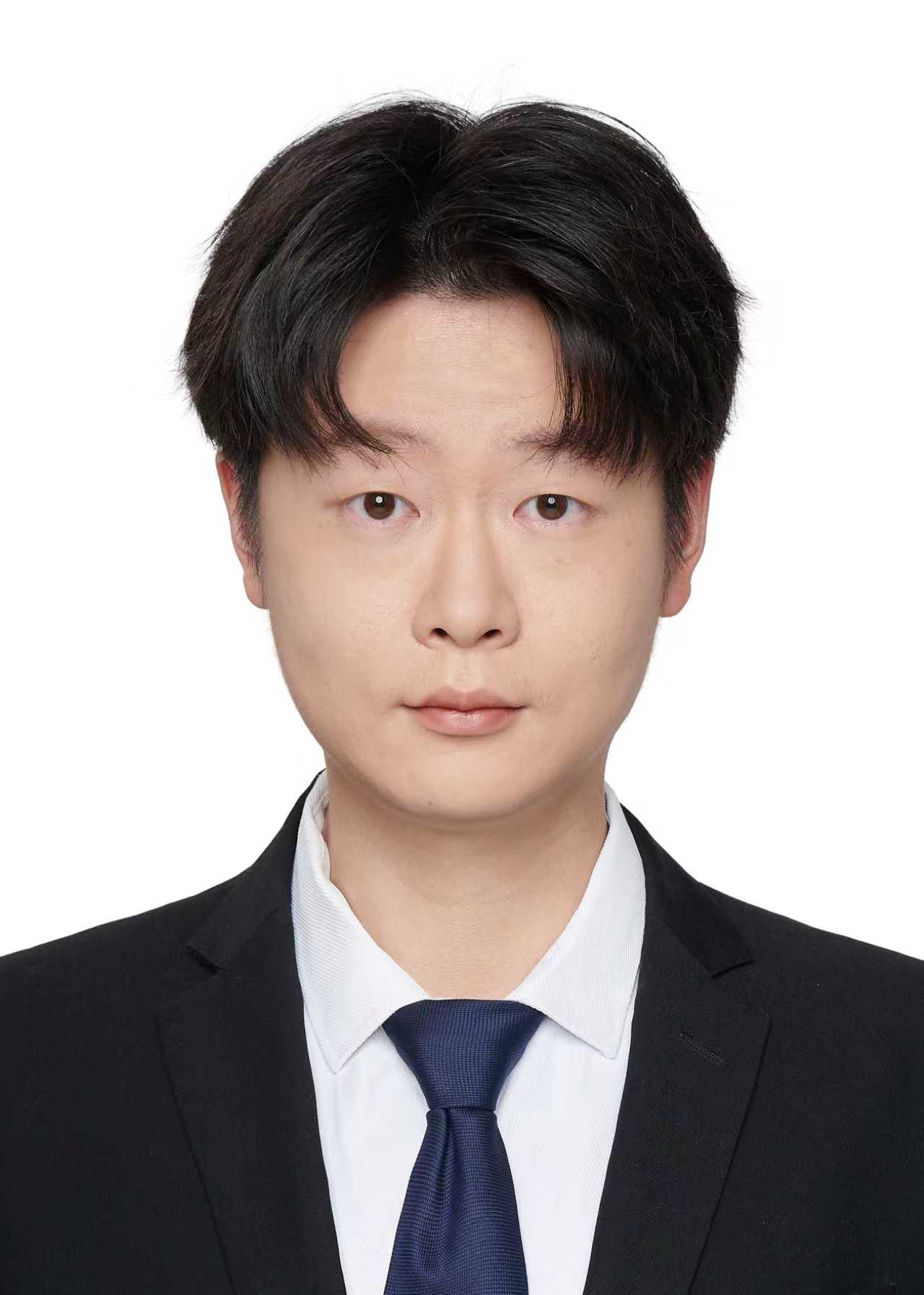}}]	{Weijie Xiong} was born in Sichuan, China, in 1999. He received the B.S. degree in electronic information engineering from the University of Electronic Science and Technology of China (UESTC), Chengdu, China, in 2021, where he is currently pursuing the Ph.D. degree in information and communication engineering. His research interests include waveform design for radar and communication systems, intelligent reflecting surfaces, optimization theory, and deep learning.
\end{IEEEbiography}

\begin{IEEEbiography}[{\includegraphics[width=1in,height=1.25in,clip,keepaspectratio]{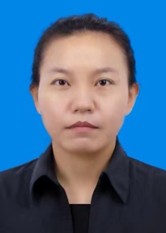}}] {Guomin Sun} received the bachelor's degree from Shanxi Normal University, Shanxi, China, in 2012, and the Ph.D. degree from the University of Electronic Science and Technology of China (UESTC), Sichuan, China, in 2020. She is currently an Assistant Research Fellow with the School of Information and Communication Engineering, UESTC. Her research interests include deep learning and RF transmitter identification.
\end{IEEEbiography}

\begin{IEEEbiography}[{\includegraphics[width=1in,height=1.25in,clip,keepaspectratio]{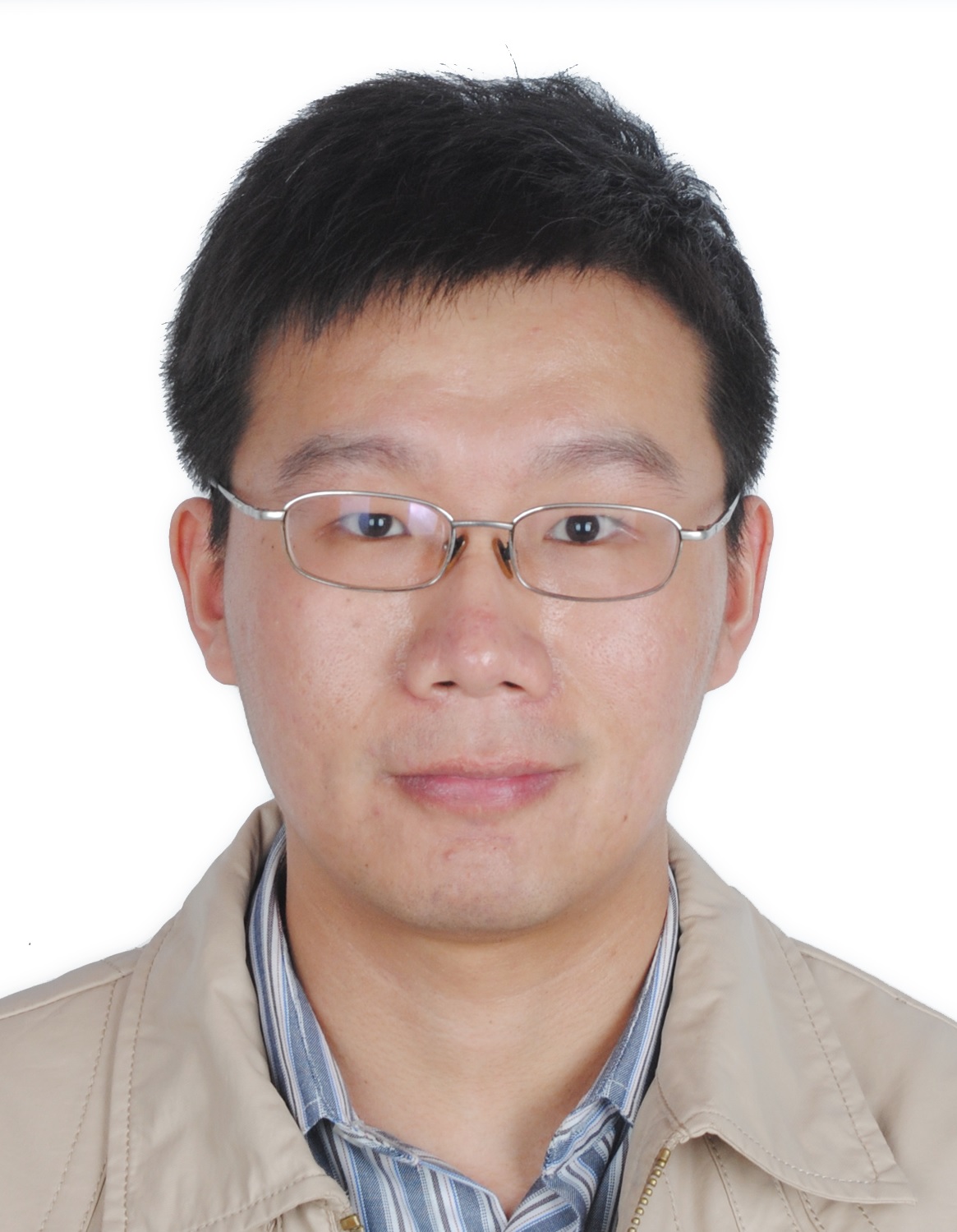}}] {Jingran Lin}  received the B.S. degree in Computer Communication from University of Electronic Science and Technology of China (UESTC), Chengdu, China, in 2001, and the M.S. and Ph.D. degrees in Signal and Information Processing from UESTC in 2005 and 2007, respectively.	After his graduation in June 2007, he joined the School of Information and Communication Engineering, UESTC, where he is currently a Full Professor. From January 2012 to January 2013, he was a Visiting Scholar with the University of Minnesota (Twin Cities), Minneapolis, MN, USA. His research interests include the algorithm design and analysis for the intelligent signal processing problems arising from modern communication systems.
\end{IEEEbiography}

\vfill

\end{document}